%% file: paper_homhbfem_revised_2.tex
\documentclass[%
 reprint,
 amsmath,amssymb,
 aps,
]{revtex4-2}

\usepackage{graphicx}% Include figure files
\usepackage{dcolumn}% Align table columns on decimal point
\usepackage{bm}% bold math
\usepackage{amssymb}
\usepackage{amsmath}
\usepackage{bm}
\usepackage{mathtools}
\usepackage{tikz}
\usepackage{pgfplots}
\usepackage{siunitx}
\usepackage{caption}
\usepackage{booktabs}
\usepackage{multirow}
\usepackage{standalone}
\usepackage{float}
\usepackage{subcaption}
\usepackage{nicefrac}
\usepackage{xspace}
\usepackage{graphicx}
\usepackage{textcomp} % for trademark symbol
\usepackage{latexsym} % for trademark symbol
\usetikzlibrary{shapes.geometric, arrows, positioning,shapes.multipart}
\definecolor{verylightblue}{RGB}{0,156,218}
\definecolor{verylightpink}{RGB}{201,48,142}
\definecolor{red}{RGB}{230,0,26}
\definecolor{green}{RGB}{153,192,0}
\definecolor{blue}{RGB}{0,90,169}
\definecolor{lightblue}{RGB}{0,131,204}
\definecolor{lightsmaragd}{RGB}{0,131,204}
\definecolor{orange}{RGB}{236,101,0}
\definecolor{lightpurple}{RGB}{166,0,132}
\definecolor{darkblue_one}{RGB}{0,78,138} % 1c
\definecolor{darkblue_two}{RGB}{0,104,157} % 2c
\definecolor{darkemerald}{RGB}{0,136,119} % 3c
\definecolor{darkgreen_one}{RGB}{127,171,22} % 4c
\definecolor{darkgreen}{RGB}{127,171,22} %4c

\definecolor{darkgreen_two}{RGB}{177,189,0} % 5c
\definecolor{darkyellow_one}{RGB}{215,172,0} % 6c
\definecolor{darkyellow_two}{RGB}{210,135,0} % 7c
\definecolor{darkyellow}{RGB}{210,135,0} % 7c

\definecolor{darkorange}{RGB}{204,76,3} % 8c

\definecolor{darkred}{RGB}{185,15,34} % 9c

\definecolor{darkpurple}{RGB}{149,17,105} % 10c

\tikzstyle{start} = [rectangle, rounded corners, minimum width=4.2cm, minimum height=0.5cm, text centered, draw=darkred, line width = 0.6mm]
\tikzstyle{stop} = [rectangle, rounded corners, minimum width=4.2cm, minimum height=0.5cm, text centered, draw=darkblue_two, line width = 0.6mm]
\tikzstyle{process_blue} = [rectangle, minimum width=4.2cm, minimum height=1cm, text centered, draw=darkblue_two, line width = 0.6mm]
\tikzstyle{process_red} = [rectangle, minimum width=4.2cm, minimum height=1cm, text centered, draw=darkred, line width = 0.6mm]
\tikzstyle{process_black} = [rectangle, minimum width=4.2cm, minimum height=1cm, text centered, draw=black, line width = 0.6mm]
\tikzstyle{decision} = [diamond, aspect=4, minimum width=3cm, minimum height=0.1cm, text centered, draw=darkblue_two, line width = 0.6mm]
\tikzstyle{arrow} = [thick,->,>=stealth]

\renewcommand{\epsilon}{\varepsilon}

\newcommand{\nutensor}{\ensuremath{\overline{\overline{\nu}}}}
\newcommand{\sigmatensor}{\ensuremath{\overline{\overline{\sigma}}}}

\renewcommand{\d}{\ensuremath{ \mathrm{d}} }

\newcommand{\AvecComplex}{\ensuremath{\underline{\vec{A}}}}

\newcommand{\Hvc}{\ensuremath{\Vec{\underline{H}}}}

\newcommand{\JsvecComplex}{\ensuremath{\underline{\vec{J}}_{\mathrm{s}}}}

\newcommand{\LossPerpTimeAv}{\ensuremath{\overline{P}_{\perp}}}
\newcommand{\LossParTimeAv}{\ensuremath{\overline{P}_{\parallel}}}
\newcommand{\OmegalsHom}{\widetilde{\Omega}_{\text{ls}}} % homogenized lamination stack

\newcommand{\DirichletBd}{\Gamma_{\mathrm{D}}}

\newcommand{\dV}{\ensuremath{ \mathrm{d}V}}

\newcommand{\nuComp}{\ensuremath{\underline{\nu}}}

\newcommand{\K}{\ensuremath{\mathbf{K}}} % generic stiffness matrix
\newcommand{\KnuTwf}[1]{\ensuremath{\mathbf{K}_{\underline{\nutensor}_0 \left(#1 \omega_{\mathrm{f}} \right)}}} % stiffness matrix with reluctivity tensor evaluated at given harmonic of fundamental frequency
\newcommand{\MsigmaT}{\ensuremath{\mathbf{M}_{\sigmatensor}}}
\newcommand{\Knuh}[1]{\ensuremath{\K_{\underline{\nu}_{#1}}}}

\newcommand{\adjustedaccent}[1]{%
	\mathchoice{}{}
	{\mbox{\raisebox{-.5ex}[0pt][0pt]{$\scriptstyle#1$}}}
	{\mbox{\raisebox{-.35ex}[0pt][0pt]{$\scriptscriptstyle#1$}}}
}

\newcommand{\adjustedaccentsecondbow}[1]{%
	\mathchoice{}{}
	{\mbox{\raisebox{-.7ex}[0pt][0pt]{$\scriptstyle#1$}}}
	{\mbox{\raisebox{-.35ex}[0pt][0pt]{$\scriptscriptstyle#1$}}}
}

\newcommand\bow[1]{\overset{\adjustedaccent{\smallfrown}}{#1}}
\newcommand\secondbow[1]{\overset{\adjustedaccentsecondbow{\smallfrown}}{#1}}

\newcommand{\ahatc}[1]{\ensuremath{\bow{\underline{\mathbf{a}}}_{#1}}} % magnetic vector potential (complex)
\newcommand{\ff}{\ensuremath{f_{\mathrm{f}}}}
\newcommand{\wf}{\ensuremath{\omega_{\mathrm{f}}}}

\begin{document}
\mbox{Published in Phys.\ Rev.\ Accel.\ Beams \textbf{28},\ 104601\ (2025), doi:\ 10.1103/9mnn-w7lj }
\preprint{APS/123-QED}

\title{Homogenized harmonic balance finite element method for\\ nonlinear eddy current simulations of fast corrector magnets}% Force line breaks with \\

\author{Jan-Magnus Christmann}
\email{jan-magnus.christmann@tu-darmstadt.de}
\author{Laura A. M. D'Angelo }%

 \author{Herbert De Gersem}
\affiliation{%
 Institute for Accelerator Science and Electromagnetic Fields, Technical University of Darmstadt, Germany\\
}%


\author{Sven Pfeiffer}
\author{Sajjad H. Mirza}
\author{Matthias Thede}
\author{Alexander Aloev}
\author{Holger Schlarb}
\affiliation{
 Deutsches Elektronen-Synchrotron DESY, Hamburg, Germany\\
}%
%\affiliation{
%\affiliation{%



\begin{abstract}
This paper develops a homogenized harmonic balance finite element method (HomHBFEM) to predict the dynamic behavior of magnets with fast excitation cycles, including eddy current and skin effects. A homogenization technique for laminated yokes avoids resolving the individual laminates and the skin depth in the finite element (FE) mesh. Instead, the yoke is represented by a bulk surrogate material with frequency-dependent parameters. The ferromagnetic saturation of the yoke at higher excitation currents is tackled by a harmonic balance method, which accounts for a coupled set of frequency components. Thereby, a computationally expensive time-stepping of the eddy-current field problem and a convolution of the homogenized yoke model are avoided. The HomHBFEM enables, for the first time, to conduct nonlinear simulations of fast corrector magnets, which are embedded in a fast orbit feedback system to counteract orbit disturbances over a broad frequency spectrum, and thus guarantee a stable light-source operation. The results show the impact of the nonlinearity on the phase lag and the field attenuation as well as the eddy current losses at frequencies up to several tens of kilohertz. The numerical validation for a C-dipole magnet example shows that the HomHBFEM achieves a sufficient accuracy at an affordable computational effort, with simulation times of a few hours. In comparison, standard 3D transient FE simulations need to resolve the lamination thickness and the skin depth in space and the largest relevant frequency in time, which leads to a two to three orders of magnitude larger mesh and prohibitive computational effort, with simulation times of a few weeks on a contemporary computer server.

%\begin{description
%\end{description}
\end{abstract}

\maketitle

%\tableofcontents

\section{Introduction}\label{sec:intro} 

% Move 1: Establishing a territory
The transition to the fourth generation of storage ring based synchrotron radiation sources is in full swing, with accelerator laboratories worldwide upgrading their facilities~\cite{shin_2021}. At the core of these upgrades is the implementation of the multibend achromat (MBA) design, or variants thereof, for the magnetic lattice~\cite{einfeld_1995}. This allows for a significant reduction in the electron beam's emittance and ultimately for higher brightness of the photon beam. In the case of {PETRA IV}, the fourth-generation storage ring currently being planned at DESY, the aim is to reach a horizontal emittance of  $\SI{20}{\pico\meter\radian}$ and a vertical emittance of $\SI{4}{\pico\meter\radian}$~\cite{petraiv_tdr,mirza_2023}. With the push towards such extremely low emittances comes the challenge of providing the necessary electron beam stability \cite{winick_1997}. Typically, the requirement is to stabilize the beam position within $\SI{10}{\percent}$ of the beam size \cite{tian_2015, chiu_2016}. To that end, fast orbit feedback systems (FOFB) are employed. These control systems employ beam position monitors to measure the perturbation of the orbit and fast corrector (FC) magnets to steer the particles toward the design orbit \cite{uzun_2005}. For {PETRA~IV}, the FOFB should achieve a disturbance-rejection bandwidth of $\SI{1}{\kilo\hertz}$~\cite{mirza_2023}. To meet the space constraints, the FC magnets are also used for slow correction. Hence, we need to understand the behavior of the FC magnets from the DC case up to the kilohertz range. To that end, finite element (FE) simulations of the eddy current effects in the laminated yokes of the FC magnets must be conducted. Since the magnets are short compared to their transversal dimensions, a three-dimensional (3D) analysis is indispensable.
 
It is well known that such 3D FE simulations of laminated ferromagnetic structures at elevated frequencies are computationally exceedingly demanding~\cite{gyselinck_2015}. Reaching convergence becomes more and more difficult at higher frequencies since the thin lamination sheets combined with the small skin depth necessitate an extremely fine mesh. As a result, a brute-force approach, i.e., simulating the laminated structures using commercial software tools with a very fine mesh, is not an option for realistic models~\cite{sabariego_2020}. For simulations with a linear \mbox{$B$-$H$ curve}, there are appropriate homogenization techniques that allow to capture the eddy current effects without resolving the laminations and the skin depth with the FE mesh~\cite{dular3DMagneticVector2003}. We have employed such a technique to conduct linear simulation studies of the FC~magnets for PETRA IV in~\cite{christmannFiniteElementSimulation2024a} and~\cite{christmann_findings}. However, if a nonlinear \mbox{$B$-$H$ curve} is to be considered, the only existing option is to resort to multiscale methods, see \mbox{e.g.~\cite{niyonzima_2013,niyonzima_2016,dular_2008,hollaus_2019}}, which are generally too complicated and therefore not appropriate to conduct extensive simulation studies of larger 3D models like the FC magnets. 

To enable nonlinear eddy current simulations of FC magnets, we combine a relatively simple homogenization technique \cite{dular3DMagneticVector2003} with the harmonic balance finite element method (HBFEM) \cite{YamadaBessho1988}. The homogenization technique replaces the laminations with a bulk model characterized by frequency-dependent material tensors and the HBFEM introduces a multi-harmonic approach, i.e., the solution is represented as a truncated Fourier series. Then, we can simulate the FC magnets with a relatively coarse mesh and we can include a nonlinear $B$-$H$ curve without costly time-stepping. 

This paper is structured as follows. First, in Sec.~\ref{sec:hom} and Sec.~\ref{sec:hbfem}, we briefly explain the homogenization technique and the HBFEM individually. Then, in Sec.~\ref{sec:homhbfem}, we explain how the HBFEM and the homogenization technique are combined into the so-called homogenized harmonic balance finite element method (HomHBFEM). In Sec.~\ref{sec:toy_model} and Sec.~\ref{sec:c_magnet}, the HomHBFEM is verified by comparing results for a model of a laminated inductor and a C-dipole magnet to results obtained by 3D transient FE simulations with an adequately fine mesh carried out in CST Studio Suite\textsuperscript{\textregistered}~\cite{CST}. The verification studies clearly show the significant reduction in required computational resources and simulation times achieved with the HomHBFEM. Next, in Sec.~\ref{sec:corrector}, the method is applied to a model of the FC magnets for {PETRA IV} and the results are discussed. Finally, conclusions are formulated in Sec.~\ref{sec:conclusion}.

\section{Homogenization}\label{sec:hom} 
The homogenization technique considered here consists in replacing a lamination stack with a frequency-dependent anisotropic surrogate model~\cite{dular3DMagneticVector2003}. We have successfully applied the homogenization to the laminated yokes in linear models of the FC magnets for \mbox{PETRA IV} in~\cite{christmannFiniteElementSimulation2024a} and~\cite{christmann_findings}. In these contributions, we have investigated the eddy current losses, magnetic flux densities along the axis, and the integrated transfer function of the magnets by conducting (linear) frequency domain simulations. Cross-talk with the neighboring quadrupole magnets has also been considered~\cite{christmann_findings}. In the following, we will give a short explanation of the technique, for the derivation, we refer the reader to~\cite{dular3DMagneticVector2003}. 

Let $\Omega$ be the computational domain and let $H_{\mathrm{D}}\left(\text{curl};\Omega \right)$ be the Sobolev space consisting of all square-integrable vector fields $\underline{\vec{v}} : \Omega \to \mathbb{C}^{3}$ whose (weak) curl is square-integrable and whose tangential components vanish on the Dirichlet part of the boundary, i.e., 
\begin{multline}
	H_{\mathrm{D}}\left(\text{curl}; \Omega \right) \coloneqq \{ \underline{\vec{v}} \in L^2 ( \Omega;\mathbb{C}^{3}) : \nabla \times \underline{\vec{v}} \in L^2 ( \Omega;\mathbb{C}^{3}), \\ \vec{n} \times \underline{\vec{v}} \vert_{\DirichletBd} = 0   \}.
\end{multline} 
Then, the frequency domain representation of the weak formulation of the magnetoquasistatic problem which we will use throughout this work reads:
\mbox{Determine ${\AvecComplex \in H_{\mathrm{D}}\left(\text{curl};\Omega \right)}$ such that}
\begin{multline}\label{eq:WF1}
	\int_{\Omega} \left( \nu \nabla \times \AvecComplex\right) \cdot \left(\nabla \times \AvecComplex' \right)\,\d V + \jmath \omega \int_{\Omega} \sigma \AvecComplex \cdot \AvecComplex'\,\d V = \\ \int_{\Omega} \JsvecComplex \cdot \AvecComplex'\,\d V \, \, \, \, \, \forall \AvecComplex' \in  H_{\mathrm{D}}\left(\text{curl}; \Omega \right),
\end{multline}
where $\omega$ is the angular frequency, the $\AvecComplex'$ are the test functions, $\AvecComplex$ is the magnetic vector potential, $\JsvecComplex$ the source current density, $\sigma$ the conductivity, and $\nu$ the reluctivity.  The weak formulation in Eq.~\eqref{eq:WF1} is called $\vec{A}^{*}$-formulation~\cite{emson_1983} and it is one of many options to state the magnetoquasistatic problem~\cite{meunier_2008}. Herein, the magnetic flux density $\vec{\underline{B}}$ is given by $\vec{\underline{B}} = \nabla \times \AvecComplex$ and the magnetic field strength $\vec{\underline{H}}$ is obtained via $\vec{\underline{H}} = \nu \vec{\underline{B}}$.

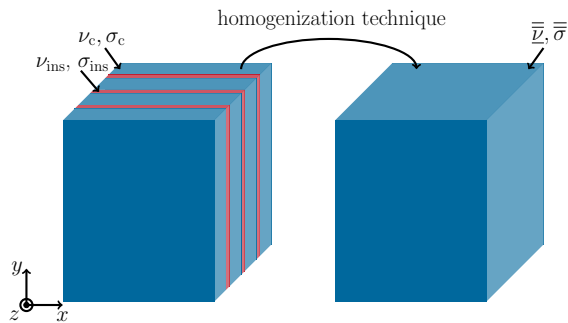
\begin{figure}[htbp]
	\centering
	\scalebox{0.4}{\input{fig_1.tex}} 
	\caption{Left: Lamination stack with insulation in red and conducting laminates in blue. Right: Homogenized model.}
	\label{fig:homogenization}
\end{figure}

Inside of a lamination stack, $\sigma(\vec{r})$ and $\nu(\vec{r})$ are functions of the spatial coordinate $\vec{r}$, since they are different for the conducting laminates and insulation sheets. Let the coordinate system be chosen as shown in Fig.~\ref{fig:homogenization}. Then, the homogenization technique consists in replacing $\sigma(\vec{r})$ and $\nu(\vec{r})$ in the yoke with the spatially constant material tensors
\begin{align}
	\sigmatensor &= \sigma_{\mathrm{c}} \begin{bmatrix}
		1 & 0 & 0 \\ 0 & 1 & 0 \\ 0 & 0 & 0
	\end{bmatrix}, \label{eq:tensor_conductivity} \\
	\begin{split}
		\underline{\nutensor} &=  \frac{\sigma_{\mathrm{c}} d \delta \omega \left(1+\jmath\right)}{8} \frac{\sinh \left( \left(1 + \jmath \right) \frac{d}{\delta}\right)}{\sinh^2\left( \left(1 + \jmath \right) \frac{d}{2\delta}\right)} 
		\begin{bmatrix}
			1 & 0 & 0 \\ 0 & 1 & 0 \\ 0 & 0 & 0 
		\end{bmatrix} \\  & +  \nu_{\mathrm{c}} \begin{bmatrix}
			0 & 0 & 0 \\ 0 & 0 & 0 \\ 0 & 0 & 1
		\end{bmatrix}, \label{eq:tensor_reluctivity}
	\end{split}
\end{align}
where $\sigma_{\mathrm{c}}$ denotes the conductivity of the laminates, $\nu_{\mathrm{c}}$ their reluctivity, $d$ their thickness, and $ \delta = \sqrt{\frac{2 \nu_{\mathrm{c}}}{\sigma_{\mathrm{c}}\omega}}$ is the skin depth.

Equations \eqref{eq:tensor_conductivity} and \eqref{eq:tensor_reluctivity} assume that the 
insulation thickness is negligible compared to the lamination thickness. To consider a non-negligible insulation thickness, we adapt the technique slightly using the stacking factor $\gamma$ which is the percentage of the yoke's volume consisting of conducting material. Regarding the conductivity tensor, we simply multiply the $x$- and $y$-component with $\gamma$, while the $z$-component remains zero. For the reluctivity tensor we use the modified in-plane components
\begin{equation}\label{eq:hom_adapted_1}
	\underline{\nu}_{xy}^{\mathrm{mod}} = \frac{1}{\left( 1 - \gamma \right) \nu_{\mathrm{ins}}^{-1} + \gamma \underline{\nu}_{xy}^{-1} }
\end{equation}
and the modified perpendicular component
\begin{equation}\label{eq:hom_adapted_2}
	\underline{\nu}_{z}^{\mathrm{mod}} = \gamma \nu_z + \left(1 - \gamma\right)\nu_{\mathrm{ins}},
\end{equation}
where $\nu_{\mathrm{ins}}$ is the reluctivity of the insulation, $\underline{\nu}_{xy}$ denotes the $x$- and $y$-components of the reluctivity tensor in Eq.~\eqref{eq:tensor_reluctivity} and $\nu_z$ denotes its $z$-component. These modifications of the conductivity and reluctivity tensors using the stacking factor $\gamma$ are based on a widespread technique that has for example been used to simulate the bending magnets of the SIS100 synchrotron at FAIR, Darmstadt, Germany~\cite{koch_2008}. For detailed information on this technique, see for example~\cite{degersem_2012,silva_1995,kaimori_2007}. 

The basic idea behind the choice of the averaged reluctivities is that in the $z$-direction, the magnetic flux has to pass a series connection of magnetic resistances, whereas in the $x$- and $y$-direction, the magnetic flux traverses a parallel connection of magnetic resistances. For the choice of the conductivity, the idea is that due to the presence of the insulation, in $x$- and $y$-direction, the cross-sectional area of the current paths is reduced by the factor $\gamma$, whereas in $z$-direction, the current flow is entirely suppressed. Although in reality, local insulation faults may permit small leakage currents between lamination sheets, which could theoretically lead to increased power loss, such effects are typically negligible. Provided that appropriate care is taken to avoid short circuits of multiple laminations, e.g., by clamping elements or excessive burrs, the assumption of zero interlaminar current flow is thus valid~\cite{handgruber_2013, marion_1995}. 

It should be noted here that there are multiple other homogenization techniques following the same pattern, thus introducing particular choices of $\sigmatensor$ and $\underline{\nutensor}$, that are different from the one in Eqs.~\eqref{eq:tensor_conductivity} and \eqref{eq:tensor_reluctivity}, see for example~\cite{wang_new_2011,gyselinckCalculationEddyCurrents1999a}. Despite our focus on the choices given by Eqs.~\eqref{eq:tensor_conductivity} and~\eqref{eq:tensor_reluctivity}, any other homogenization technique fits within the framework of the HomHBFEM described below. For instance, if the laminates are relatively thick, it might be advisable to use a conductivity tensor with a non-zero $z$-component, not to model interlaminar current flow, but to better capture averaged eddy current behavior, as suggested in~\cite{wang_new_2011} and~\cite{hahne_1996}.

\section{Harmonic Balance FEM}\label{sec:hbfem}
The core idea of the harmonic balance method is to approximate the solution of a nonlinear time periodic differential equation with a truncated Fourier series. This approach is well known in mathematics, for example to solve periodic boundary value problems for ordinary differential equations \cite{deuflhardNewtonMethodsNonlinear2005, hayashi_2014}. 
The combination of this technique with the FEM is commonly referred to as HBFEM. For the solution of nonlinear magnetic field problems, it was first proposed in~\cite{YamadaBessho1988}. Since then, the method has been developed further, see for instance \cite{yamadaHarmonicBalanceFinite1989, yamadaCalculationNonlinearEddycurrent1991,zhaoAnalysisDCBias2011}, and has become a valuable alternative to transient simulations if one is interested in the steady-state solution of a nonlinear eddy current problem. The main advantage of the HBFEM over a transient simulation is that it does not require any time-stepping.

To derive the method, we start from the strong form of the magnetoquasistatic partial differential equation in time domain, which reads
\begin{equation}\label{eq:MQS_NL_TD}
	\nabla \times \left(\nu\left( t \right)  \nabla\times \vec{A}\left(t\right) \right) + \sigma \frac{\partial \vec{A} \left( t \right)}{\partial t} = \vec{J}_{\mathrm{s}}\left( t\right).
\end{equation}
The reluctivity $\nu(t)$ depends on the time $t$, since we take the nonlinearity of the $B$-$H$ curve into account. Note that we consider only non-hysteretic $B$-$H$ curves.
To carry Eq.~\eqref{eq:MQS_NL_TD} over into frequency domain, we apply the continuous Fourier transform given by 
\begin{equation}
	\underline{\vec{A}}(\omega) = \int_{-\infty}^{\infty} \vec{A}(t) \mathrm{e}^{-\jmath \omega t} \d t
\end{equation}
for the magnetic vector potential and likewise for all other time-dependent quantities. With this definition, transforming Eq.~\eqref{eq:MQS_NL_TD} into the frequency domain yields
\begin{equation}\label{eq:MQS_NL_FD}
	\frac{1}{2 \pi} \nabla \times  \left( \nuComp\left(\omega\right) * \nabla \times \AvecComplex \left(\omega\right) \right) + \jmath \omega \sigma \AvecComplex\left( \omega\right)  = \JsvecComplex \left( \omega \right),
\end{equation}
where $*$ denotes the convolution operator.

Since all time-dependent quantities in Eq.~\eqref{eq:MQS_NL_TD} are periodic functions of time, they have discrete spectra, i.e., the spectra have nonzero values only at multiples of the so-called fundamental frequency $\wf$. Hence, we can rewrite the Fourier transforms in Eq.~\eqref{eq:MQS_NL_FD} as sums of unit impulses~\cite{oppenheimSignalsSystems1982}, so-called Dirac combs, and obtain
\begin{align}\label{eq:dirac_combs}
	\frac{1}{2 \pi} \nabla \times \Bigg(
	& \sum_{n=-\infty}^{\infty} 2 \pi \underline{\nu}_{n} \delta \left(w - n\wf \right) * \notag \\
	& \sum_{n=-\infty}^{\infty} 2 \pi \nabla \times \underline{\vec{A}}_{n} \delta \left(w - n\wf \right)
	\Bigg) \notag \\ &+\sum_{n = -\infty}^{\infty} \jmath \sigma  n \wf 2 \pi \underline{\vec{A}}_{n} \delta \left(w - n\wf \right) \notag \\
	& = \sum_{n=-\infty}^{\infty} 2 \pi \underline{\vec{J}}_{\mathrm{s},n} \delta \left(w - n\wf \right),
\end{align}
where $\delta$ is the Dirac delta function and $ \AvecComplex_n$,  $ \underline{\Vec{J}}_{\mathrm{s},n} $, and $\nuComp_{n}$ denote the Fourier series coefficients of the respective quantity. Note that since the DC component of the reluctivity is real-valued, we will refer to it as $\nu_0$ and therefore, to avoid confusion, the vacuum reluctivity will be denoted by $\nu_{\mathrm{vac}}$. Computing the convolution of the two Dirac combs in Eq.~\eqref{eq:dirac_combs} yields~\cite{bracewell_1986}
\begin{align}
	&\sum_{n = -\infty}^{\infty} \left( \sum_{k = -\infty}^{\infty} \nabla \times \left( \underline{\nu}_{k} \nabla \times \vec{\underline{A}}_{n-k} \right) \right) \delta\left( w - n\wf\right) \notag \\ 
	&+\sum_{n = -\infty}^{\infty} \jmath \sigma  n \wf \underline{\vec{A}}_{n} \delta \left(w - n\wf \right) \notag \\
	& = \sum_{n=-\infty}^{\infty}  \underline{\vec{J}}_{\mathrm{s},n} \delta \left(w - n\wf \right).
\end{align}
Thus, for each $n \in \mathbb{Z}$, we have
\begin{equation}\label{eq:MultiharmonicPDE_StrongForm}
	\sum_{k=-\infty}^{\infty} \nabla \times \left( \nuComp_k(\AvecComplex) \nabla \times \AvecComplex_{n-k} \right) + \jmath n \wf \sigma \AvecComplex_n = \underline{\Vec{J}}_{\mathrm{s},n}.  
\end{equation}
Herein, we use the notation $\nuComp_k(\AvecComplex)$ to indicate that, since we are considering a nonlinear magnetization characteristic, each Fourier series coefficient of the reluctivity, $\nuComp_k$, depends on all Fourier series coefficients of the magnetic vector potential $\AvecComplex$. 

Based on Eq.~\eqref{eq:MultiharmonicPDE_StrongForm}, we can derive the weak formulation for each harmonic component in the standard manner, by multiplying with the test function $\Vec{\underline{A}}'$ and integrating over the computational domain. This leads to 
\begin{multline}\label{eq:MultiharmonicPDE_WeakForm}
	\int_\Omega \left( \sum_{k=-\infty}^{\infty} \nuComp_k \left(\AvecComplex \right) \nabla \times \AvecComplex_{n-k} \right) \cdot \nabla \times \underline{\Vec{A}}' \dV \\ + \jmath n \wf \sigma \int_\Omega \AvecComplex_n \cdot \underline{\Vec{A}}' \dV \\
	= \int_\Omega \underline{\Vec{J}}_{\mathrm{s},n} \cdot \AvecComplex' \dV \, \, \, \, \, \forall \AvecComplex' \in  H_{\mathrm{D}}\left(\text{curl}; \Omega \right).
\end{multline}
Discretizing Eq.~\eqref{eq:MultiharmonicPDE_WeakForm} with edge elements finally results in 
\begin{equation}\label{eq:FEM}
	\sum_{k=-\infty}^{\infty} \mathbf{K}_{\nu_k} \left(\bow{\underline{\mathbf{a}}}\right)\bow{\underline{\mathbf{a}}}_{n-k} + \jmath n\wf\mathbf{M}_{\sigma}\bow{\underline{\mathbf{a}}}_{n} = \secondbow{\bow{\underline{\mathbf{j}}}}_{\mathrm{s},n},
\end{equation}
where $\bow{\underline{\mathbf{a}}}_{n}$ is the vector gathering the degrees of freedom (DoFs) of the $n$-th harmonic of the magnetic vector potential, $\secondbow{\bow{\underline{\mathbf{j}}}}_{\mathrm{s},n}$ is the discretized $n$-th harmonic of the source current density, $\mathbf{M}_{\sigma}$ is the mass matrix, and  $\mathbf{K}_{\nu_k}$ denotes the stiffness matrix computed with the $k$-th harmonic of the reluctivity. Analogously to the continuous case, we indicate that each harmonic of the reluctivity depends on all the harmonics of the magnetic vector potential by using the notation $\mathbf{K}_{\nu_k}\left(\bow{\underline{\mathbf{a}}}\right)$.\\

In practice, it is neither possible nor necessary to take an infinite amount of harmonics into account. Hence, we must specify the maximum order of harmonics that we want to consider in our analysis and then we truncate the discrete convolution in Eq.~\eqref{eq:FEM} accordingly. Let the maximum order of considered harmonics be $m \in \mathbb{N}$. Then, we have a nonlinear system of equations which reads
\begin{equation}\label{eq:HBFEM_final}
	\sum_{k = \max\{-m,n-m\} }^{\min\{m, n+m\}} \mathbf{K}_{\nu_k} \left( \bow{\underline{\mathbf{a}}}\right)\bow{\underline{\mathbf{a}}}_{n-k} + \jmath n\wf\mathbf{M}_{\sigma}\bow{\underline{\mathbf{a}}}_{n} = \secondbow{\bow{\underline{\mathbf{j}}}}_{\mathrm{s},n}
\end{equation}
for $ n \in \mathbb{Z} \, \cap \, [-m,m]$ \cite{roppert_2019}. Considering that $\bow{\underline{\mathbf{a}}}_{-n} = \bow{\underline{\mathbf{a}}}_{n}^{*}$, we only have to solve for $\left( \bow{\underline{\mathbf{a}}}_{n} \right)_{n \in \mathbb{N}_0}$,
which we do via successive substitution. \\

The system of equations in Eq.~\eqref{eq:HBFEM_final} is further reduced in size by recognizing that a source current density that includes only odd harmonics of the fundamental component will result in a magnetic vector potential that includes only odd harmonics as well. This can be shown by realizing that $\JsvecComplex$ including only odd harmonics can be equivalently expressed by the condition 
\begin{equation}
	\vec{J}_{\mathrm{s}} \left(t  + \frac{\pi}{\wf}\right) = - \vec{J}_{\mathrm{s}} \left(t\right), \, \, \forall t.
\end{equation}
Then, the statement follows directly from the unique solvability of Eq.~\eqref{eq:MQS_NL_TD} in the space of divergence-free $H(\mathrm{curl})$-functions with appropriate boundary and initial conditions. The full proof can be found in~\cite{bachingerNumericalAnalysisNonlinear2005a}. Furthermore, since Amp\`ere's law prescribes a linear relation between $\JsvecComplex$ and the magnetic field strength $\Hvc$, it follows that $\Hvc$ also only contains odd harmonics. 
Finally, since $\Hvc$ is related to $\AvecComplex$ via  $\Hvc = \nuComp *\nabla \times \AvecComplex$, it follows that the reluctivity $\nuComp$ only includes even harmonics \cite{degersemStrongCoupledMultiharmonic2001}. 

\section{HomHBFEM}\label{sec:homhbfem}
To understand how the HBFEM is combined with the homogenization technique, we rewrite Eq.~\eqref{eq:HBFEM_final} as
\begin{align}
	\mathbf{K}_{\nu_0}\left(\bow{\underline{\mathbf{a}}}\right)\bow{\underline{\mathbf{a}}}_{n} &+ \sum_{\substack{k= \max\{-m,n-m\}, \\ k \neq 0}}^{\min\{m, n+m\}} \mathbf{K}_{\nu_k} \left( \bow{\underline{\mathbf{a}}}\right)\bow{\underline{\mathbf{a}}}_{n-k} \notag \\  &+ \jmath n\wf\mathbf{M}_{\sigma}\bow{\underline{\mathbf{a}}}_{n} = \secondbow{\bow{\underline{\mathbf{j}}}}_{\mathrm{s},n}.
\end{align}
Now, the material tensors of the homogenization technique are introduced. The introduction of the conductivity tensor $\sigmatensor$ into the mass matrix $\mathbf{M}_{\sigma}$ is straightforward. The reluctivity tensor $\underline{\nutensor}$, on the other hand, must only be used in the construction of $\mathbf{K}_{\nu_0}$, i.e., only for the stiffness matrix containing the DC component of the reluctivity. Thereby, we obtain
\begin{align}
	\mathbf{K}_{\nutensor_0 \left(n\wf \right)}\left(\bow{\underline{\mathbf{a}}}\right)\bow{\underline{\mathbf{a}}}_{n} &+ \sum_{\substack{k= \max\{-m,n-m\}, \\ k \neq 0}}^{\min\{m, n+m\}} \mathbf{K}_{\nu_k} \left( \bow{\underline{\mathbf{a}}}\right)\bow{\underline{\mathbf{a}}}_{n-k} \notag \\ &+ \jmath n\wf\mathbf{M}_{\sigmatensor}\bow{\underline{\mathbf{a}}}_{n} = \secondbow{\bow{\underline{\mathbf{j}}}}_{\mathrm{s},n},
\end{align}
where  $\mathbf{M}_{\sigmatensor}$ is the mass matrix constructed with the conductivity tensor $\sigmatensor$ as given in Eq.~\eqref{eq:tensor_conductivity} and $\mathbf{K}_{\nutensor_0 \left(n\wf \right)}$ is the stiffness matrix constructed with the reluctivity tensor given in Eq.~\eqref{eq:tensor_reluctivity} evaluated at $w = n \wf$ and $\nu = \nu_0$ for each element, i.e., with 
\begin{align}\label{eq:homhbfem_nu_tensor}
	&\underline{\nutensor}_0\left(n \wf \right) = \notag \\ &\frac{\sigma_{\mathrm{c}} d \delta \left(n\wf, \nu_0 \right) n \wf \left(1+\jmath \right)}{8} \frac{\sinh \left( \frac{\left(1 + \jmath \right) d}{\delta\left(n\wf, \nu_0 \right) }\right)}{\sinh^2\left(  \frac{\left(1 + \jmath \right)d}{2\delta\left(n\wf, \nu_0 \right) }\right)} 
	\begin{bmatrix}
		1 & 0 & 0 \\ 0 & 1 & 0 \\ 0 & 0 & 0 
	\end{bmatrix} \notag \\  & +  \nu_{\mathrm{0}} \begin{bmatrix}
		0 & 0 & 0 \\ 0 & 0 & 0 \\ 0 & 0 & 1
	\end{bmatrix},
\end{align}
where  $\delta\left(n\wf, \nu_0 \right) = \sqrt{\frac{2 \nu_0}{n \wf \sigma_\mathrm{c}}}$. Note that the adaptations according to Eqs.~\eqref{eq:hom_adapted_1} and~\eqref{eq:hom_adapted_2} can be easily applied to the components of the tensor in Eq.~\eqref{eq:homhbfem_nu_tensor}.

To explain the implementation of the method, let us consider the following scenario. Assume that the source current contains only a first and a third harmonic and that we set the highest order of harmonics to be considered to be $m=3$. As explained in Sec.~\ref{sec:hbfem}, the fact that the source current only contains odd harmonics results in the magnetic vector potential containing only odd harmonics as well and the reluctivity containing only even harmonics. Hence, we arrive at the following nonlinear system of equations
\begin{subequations}\label{eq:HomHbfem_simple}
	\begin{equation}
		\left( \KnuTwf{3} + \jmath 3\wf\MsigmaT \right) \bow{\underline{\mathbf{a}}}_3 + \Knuh{2} \bow{\underline{\mathbf{a}}}_1 = \secondbow{\bow{\underline{\mathbf{j}}}}_{\mathrm{s},3}, \label{eq:HomHbfem_simple_1}
	\end{equation}
	\begin{equation}
		\K_{\underline{\nu}^{*}_2} \ahatc{3} + \left(\KnuTwf{} + \jmath \wf \MsigmaT \right)\ahatc{1}  + \Knuh{2}\ahatc{1}^{*} = \secondbow{\bow{\underline{\mathbf{j}}}}_{\mathrm{s},1}.\label{eq:HomHbfem_simple_2}
	\end{equation}
\end{subequations}
Note that, for brevity, we no longer explicitly indicate the dependence of the stiffness matrices on $\bow{\underline{\mathbf{a}}}$. Linearizing the system given by Eqs.~\eqref{eq:HomHbfem_simple_1} and \eqref{eq:HomHbfem_simple_2} with the Newton-Raphson method is difficult since we are dealing with a non-analytical complex system of equations~\cite{lederer_1997,degersemStrongCoupledMultiharmonic2001}. Hence, the system is linearized using successive substitution, leading to
\begin{subequations}
	\begin{equation}
		\left( \KnuTwf{3}^{k} + \jmath 3\wf\MsigmaT \right) \ahatc{3}^{k+1} + \Knuh{2}^{k} \ahatc{1}^{k+1} = \secondbow{\bow{\underline{\mathbf{j}}}}_{\mathrm{s},3}, 
	\end{equation}
	\begin{align}
		\K_{\underline{\nu}^{*}_2}^{k} \ahatc{3}^{k+1} + \left(\KnuTwf{}^{k} + \jmath \wf \MsigmaT \right)&\ahatc{1}^{k+1} \notag \\  +\Knuh{2}^{k}&\ahatc{1}^{*k+1} = \secondbow{\bow{\underline{\mathbf{j}}}}_{\mathrm{s},1}, 
	\end{align}	
\end{subequations}
where the superscript $k = 0,1,2,...$ indicates the iteration number and $\mathbf{K}_{\underline{\nu}_i}^{k} = \mathbf{K}_{\underline{\nu}_i} ( \bow{\underline{\mathbf{a}}}^{k} )$.
In each iteration, the linearized system can now be solved as a strongly coupled block system of equations, e.g.~by a dedicated Krylov subspace solver~\cite{DeGersem_2001ah} or multigrid technique~\cite{bachinger_2006}. To avoid solving multiharmonic systems of equations increasing in size according to the number of considered harmonics, we adopt a block Jacobi iteration~\cite{saad_2003} according to
\begin{subequations}
	\begin{equation}
		\left( \KnuTwf{3}^{k} + \jmath 3\wf\MsigmaT \right) \ahatc{3}^{k+1} = \secondbow{\bow{\underline{\mathbf{j}}}}_{\mathrm{s},3}  -  \Knuh{2}^{k} \ahatc{1}^{k}, \label{eq:HomHbfem_efficient_1}
	\end{equation}
	\begin{align}
		\left(\KnuTwf{}^{k} + \jmath \wf \MsigmaT \right)\ahatc{1}^{k+1} = \secondbow{\bow{\underline{\mathbf{j}}}}_{\mathrm{s},1} -\K_{\underline{\nu}^{*}_2}^{k} &\ahatc{3}^{k} \notag\\ -&\Knuh{2}^{k}\ahatc{1}^{*k} \label{eq:HomHbfem_efficient_2}.
	\end{align}
\end{subequations}
In this way, the method is easily parallelized, i.e., the equations for the different harmonics can be solved in parallel in each iteration~\cite{yamadaCalculationNonlinearEddycurrent1991}. This is particularly important if the model is driven into strong saturation, as the number of harmonics required for accurate representation increases with the level of saturation~\cite{yao_2016}.

To compute the stiffness matrices $\mathbf{K}_{\underline{\nu}_i}^{k}$ in the $k$-th iteration, we first need to compute the discretized magnetic flux densities $\secondbow{\bow{\underline{\mathbf{b}}}}_{i}^{\raisebox{-5pt}{$\scriptstyle k$}}$ as the discrete curl of the $\bow{\underline{\mathbf{a}}}_{i}^{k}$. These complex-valued magnetic flux densities are transformed into the time domain. The time signal for the magnitude of the magnetic flux density is then inserted into the nonlinear $B$-$H$ curve to obtain the time signal for the magnitude of the magnetic field strength. By forming the quotient of these two time signals, we obtain the reluctivity~$\nu^{k}(t)$. 

Finally, computing the fast Fourier transform (FFT) of $\nu^{k}(t)$ gives us the Fourier series coefficients of the reluctivity~\cite{smith_2007}, which allows us to set up the $\mathbf{K}_{\underline{\nu}_i}^{k}$ and to assemble the system of equations given by Eqs.~\eqref{eq:HomHbfem_efficient_1} and~\eqref{eq:HomHbfem_efficient_2}. 

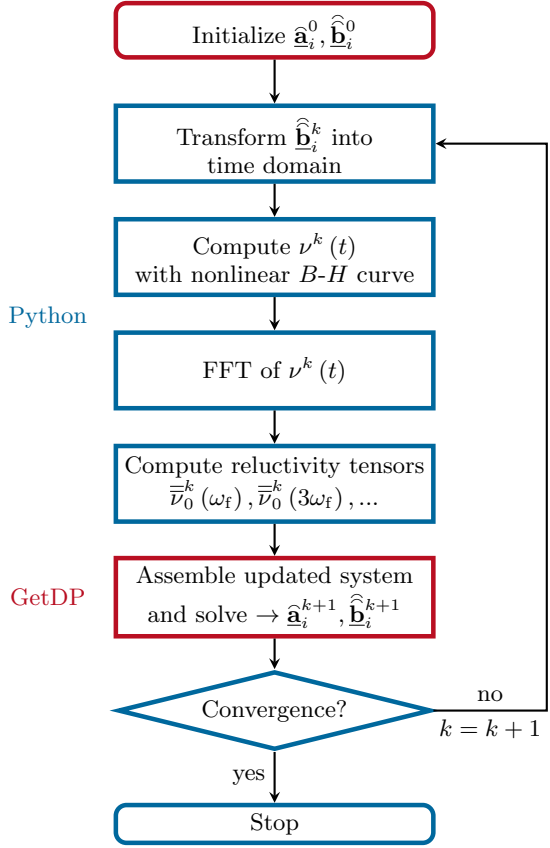
\begin{figure}[h]
	\centering
	\begin{tikzpicture}[node distance=1.5cm]
		\node (start) [start] {Initialize $\ahatc{i}^{0}, \secondbow{\bow{\underline{\mathbf{b}}}}_{i}^{\raisebox{-5pt}{$\scriptstyle 0$}}$};
		\node (pro1) [process_blue, below of=start,align=center] {Transform  $\secondbow{\bow{\underline{\mathbf{b}}}}_{i}^{\raisebox{-5pt}{$\scriptstyle k$}}$ into\\ time domain};
		\node (pro2) [process_blue, below of=pro1,align=center] {Compute $\nu^{k}\left(t\right)$\\with nonlinear $B$-$H$ curve};
		\node (pro3) [process_blue, below of=pro2,align=center] {FFT of $\nu^{k}\left( t\right)$};
		\node (pro4) [process_blue, below of=pro3,align=center] {Compute reluctivity tensors \\ $\nutensor_{0}^{k}\left(\wf\right), \nutensor_{0}^{k}\left(3\wf\right),...$};
		\node (pro5) [process_red, below of=pro4,align=center] {Assemble updated system\\ and solve $\rightarrow \ahatc{i}^{k+1}, \secondbow{\bow{\underline{\mathbf{b}}}}_{i}^{\raisebox{-5pt}{$\scriptstyle k + 1$}}$};
		\node (dec1) [decision, below of=pro5,node distance=1.5cm] {Convergence?};
		\node (stop) [stop, below of=dec1,node distance=1.5cm] {Stop};
		
		\draw [arrow] (start) -- (pro1);
		\draw [arrow] (pro1) -- (pro2);
		\draw [arrow] (pro2) -- (pro3);
		\draw [arrow] (pro3) -- (pro4);
		\draw [arrow] (pro4) -- (pro5);
		\draw [arrow] (pro5) -- (dec1);
		\draw [arrow] (dec1) -- node[anchor=east] {yes} (stop);
		\draw [arrow] (dec1.east) -- node[anchor=south] {no} node[anchor=north] {$k = k+1$}++(1.5,0) |- (pro1.east);
		% Add "python" label 
		\coordinate (python_label) at ([yshift=-0.8cm, xshift=-3cm] pro2);
		\node at (python_label) [text=darkblue_two] {Python};
	    % Add "getdp" label
		\coordinate (getdp_label) at ([yshift=-0cm, xshift=-3cm] pro5);
		\node at (getdp_label) [text=darkred] {GetDP};	
	\end{tikzpicture}
	\caption{Flow chart of the iterative procedure for the HomHBFEM. The index $i$ indicates the harmonic order and the superscript $k$ the iteration number. Blue boxes contain steps implemented in Python, red boxes steps in GetDP.}
	\label{fig:HomHbfemFlowChart}
\end{figure}

The iteration is stopped upon fulfillment of an energy-based convergence criterion, i.e., when the relative change in magnetic energy inside the ferromagnetic material from one iteration to the next is less than a given threshold. The described iterative procedure is illustrated in Fig.~\ref{fig:HomHbfemFlowChart}. It has been implemented in the open source FE software GetDP~\cite{getdp_1999} in combination with additional Python code.
To guarantee the convergence of the iterative scheme, relaxation is needed, i.e., we relax the solution $\ahatc{i}^{k+1}$ involving the solution from the previous iteration step according to
\begin{equation}
	\ahatc{i,\mathrm{relaxed}}^{k+1} = \alpha \ahatc{i}^{k+1} + \left(1 - \alpha\right) \ahatc{i}^{k},
\end{equation}
with a relaxation factor $\alpha \in \left(0,1\right]$. Then, we continue the iteration with the relaxed solution $\ahatc{i,\mathrm{relaxed}}^{k+1}$.

\section{Verification for a Laminated Inductor}\label{sec:toy_model}
\subsection{Model Description}
\begin{figure}[h]
	\centering
	\begin{minipage}[t]{0.2\textwidth}
		\centering
		\includegraphics[width=0.7\textwidth]{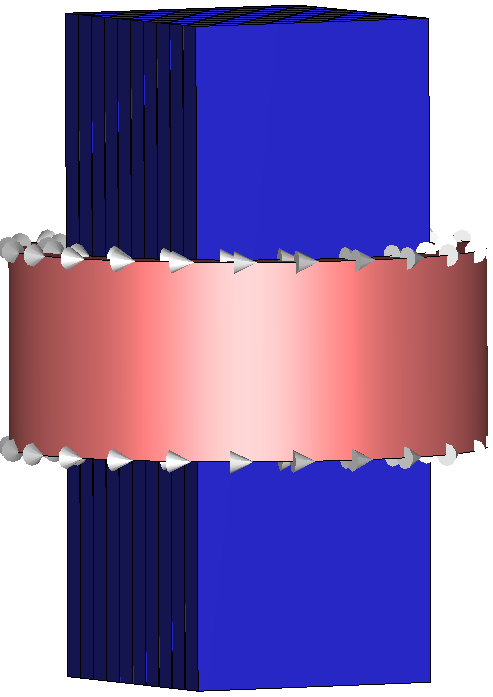}
		\vfill
		\caption{3D model of the laminated inductor.}
		\label{fig:inductor_model}
	\end{minipage}
	\hspace{0.03\textwidth} % Adjust the space between the minipages
	\begin{minipage}[t]{0.2\textwidth}
		\centering
		\input{fig_4.tex}
		\vfill
		\caption{$B$-$H$ curve based on Brauer model.}
		\label{fig:BH_curve}
	\end{minipage}
\end{figure}

To verify the HomHBFEM, we investigate the laminated inductor model shown in Fig.~\ref{fig:inductor_model}. The model consists of a coil (red) wound around a ferromagnetic core (blue) with ten laminations with a thickness of $d = \SI{0.5}{\milli\meter}$, a width of $\SI{5}{\milli\meter}$, and a height of $\SI{12.5}{\milli\meter}$. While this model does not directly resemble a FC magnet, it provides a well-suited first test case because it features laminations of a realistic thickness and it is relatively short, resulting in non-negligible stray flux perpendicular to the laminations, just like in the FC magnet. Longer models can also be simulated with the HomHBFEM but they would constitute a simpler test case, since stray flux effects become less important.

The source current $I_{\mathrm{s}}$ in the coil is chosen as 
\begin{equation}
	I_{\mathrm{s}}(t) = (\SI{1.5}{\kilo\ampere})\cos\left(\wf t\right) + (\SI{0.24}{\kilo\ampere})\cos\left(3 \wf t\right),\label{eq:excitation_current}
\end{equation}
i.e., the excitation contains the fundamental component and a third harmonic. Note that we will refer to the period of the fundamental component as $T_{\mathrm{f}}$ and to the respective frequency as $f_{\mathrm{f}}$.

For the magnetization curve, we use a modified version of the Brauer model \cite{brauerSimpleEquationsMagnetization1975}. With $H = | \vec{H} |$ and  $B = | \vec{B} |$, the Brauer curve for the nonlinear material relation is given by
\begin{equation}
	H\left( B \right) = \left( k_1 \mathrm{e}^{k_2 B^2}+ k_3 \right)B, 
\end{equation}
where $k_1,k_2,k_3 \in \mathbb{R}$ are parameters that must be fitted for given measurement data. Given the typical shape of nonlinear $B$-$H$ curves, it is clear that $k_1,k_2 > 0$.\\

The Brauer model is a popular and simple model, which has some shortcomings, one of which is the fact that the resulting reluctivity is unbounded, i.e., 
\begin{equation}
	\lim_{B \to \infty} \nu \left(B \right) = \lim_{B \to \infty}  \frac{H\left( B\right)}{B} = \infty.
\end{equation}
This is physically incorrect because in reality, the reluctivity tends towards the reluctivity of vacuum $\nu_{\mathrm{vac}}$. Furthermore, the fact that the reluctivity tends towards infinity for large $B$ is also numerically problematic, since it can lead to instabilities. Therefore, we modify the Brauer model to
\begin{align}
	\widetilde{H}(B) = 
	\begin{cases}
		( k_1 \mathrm{e}^{k_2 B^2}+ k_3 )B &, \, B \leq B_{\mathrm{s}},  \\
		( k_1 \mathrm{e}^{k_2 B_{\mathrm{s}}^2} + k_3) B_{\mathrm{s}}  + \nu_{\mathrm{vac}}\left(B - B_{\mathrm{s}} \right) &, \, B > B_{\mathrm{s}},\\
	\end{cases} 
\end{align}
where $B_\mathrm{s}$ is the saturation flux density, which we define by $\frac{\d H}{\d B} |_{B = B_{\mathrm{s}}} = \nu_{\mathrm{vac}}$. In this way, we guarantee that the modified $B$-$H$ characteristic is a $C^1$ curve and fulfills 
\begin{equation}
	\lim_{B \to \infty} \nu\left(B\right) = \lim_{B \to \infty}  \frac{\widetilde{H}\left( B\right)}{B} = \nu_{\mathrm{vac}}.
\end{equation}
A similar modification of the Brauer model is proposed in~\cite{hulsmannExtendedBrauerModel2014}. We use ${k_1 = \SI{3.8}{\meter\per\henry}}$, ${k_2 = \SI{2.17}{\tesla^{-2}}}$ and ${k_3 = \SI{396.2}{\meter\per\henry}}$, which are typical values for cold rolled steel, taken from \cite{brauerSimpleEquationsMagnetization1975}. The corresponding $B$-$H$ curve is plotted in Fig.~\ref{fig:BH_curve}. For the conductivity, we use $\sigma = \SI{10.4} {\mega\siemens\per\meter}$ and for the stacking factor, we use a realistic value of $\gamma = \SI{98.5}{\percent}$.

\subsection{Results}
We will now compare the HomHBFEM results for the laminated inductor to the results of transient nonlinear simulations conducted in CST Studio Suite\textsuperscript{\textregistered}. The HomHBFEM results have been computed with a coarse mesh that does not resolve the laminations. If not stated otherwise, field quantities up to the ninth harmonic are considered. We are using the homogenization adapted to a non-negligible insulation thickness, as described in Sec.~\ref{sec:hom}. To obtain reliable transient reference results with the CST Studio Suite\textsuperscript{\textregistered}, the mesh is chosen such that the skin depth is resolved. As a rough estimate for the skin depth, we use $\delta = \sqrt{\frac{2 \nu_{\mathrm{in}} }{\omega \sigma}}$, where $\nu_{\mathrm{in}}$ is the initial slope of the \mbox{$B$-$H$ curve}.

\subsubsection{Magnetic Energies}
First, we investigate the magnetic energies in the laminated core of the inductor. Over the course of this investigation, we will adapt the HomHBFEM in two steps, leading to three versions of the method. To distinguish between them, they will be numbered. The version of the HomHBFEM that has been introduced up until this point is called \mbox{HomHBFEM-1}. Figure~\ref{fig:tm_energy__without_trafo_pos_1} shows the results for selected frequencies ${f_{\mathrm{f}} \in \{50, 100, 500, 1000\}\,\si{Hz}}$. While we observe a good agreement for ${f_{\mathrm{f}}=\SI{50}{\hertz}}$ and ${f_{\mathrm{f}}=\SI{100}{\hertz}}$, there are significant differences between \mbox{HomHBFEM-1} and the transient reference results at ${f_{\mathrm{f}}=\SI{500}{\hertz}}$ and $f_{\mathrm{f}}={\SI{1}{\kilo\hertz}}$. At these frequencies, the energy computed from the transient reference simulation does no longer fall to zero. This is mainly due to the fact that because of the skin effect, the magnetic flux densities across one lamination are not in phase, e.g. there is a significant phase shift between the magnetic flux density in the center of a lamination and the flux density close to its surface. This phase shift increases with rising frequency.
\begin{figure}[h]
	\centering
	% First row of figures
	\begin{subfigure}{0.23\textwidth}
		\centering
		\begin{tikzpicture}[scale=0.51]
			\begin{axis}[
				xlabel = {$t\,(\si{\milli\second})$},
				xlabel style={ yshift = 5pt, font=\large},
				ylabel= {$W_{\mathrm{mag}}\,(\si{\milli\joule})$},
				ylabel style={ yshift = -10pt,font=\large},
				yticklabel style={/pgf/number format/fixed},
				grid = both,
				minor grid style = {gray!15},
				legend style={at={(0.05,0.12)},anchor = west,nodes={scale=0.9, transform shape}},
				y filter/.code={\pgfmathparse{1000*\pgfmathresult}},
				x filter/.code={\pgfmathparse{1000*\pgfmathresult}},
				xmin = -2,
				xmax = 82,
				ymin = -0.02,
				ymax = 0.34,
				]
				\addplot[line width = 1.5pt, color=darkgreen] table {data_fig_5a_homhbfem.txt};
				\addlegendentry{HomHBFEM-1};
				\addplot[line width = 1.5pt, color=darkred,dashed] table {data_fig_5a_cst.txt};
				\addlegendentry{reference result};
			\end{axis}
		\end{tikzpicture}
		\caption{$f_{\mathrm{f}} = \SI{50}{\hertz}$.}
		\label{fig:tm_energy_50Hz}
	\end{subfigure}
	\hspace{0.1cm}
	\begin{subfigure}{0.23\textwidth}
		\centering
		\begin{tikzpicture}[scale=0.51]
			\begin{axis}[
				xlabel = {$t\,(\si{\milli\second})$},
				xlabel style={ yshift = 5pt, font=\large },
				ylabel= {$W_{\mathrm{mag}}\,(\si{\milli\joule})$},
				ylabel style={ yshift = -10pt, font=\large},
				yticklabel style={/pgf/number format/fixed},
				grid = both,
				minor grid style = {gray!15},
				legend style={at={(0.05,0.1)},anchor = west,nodes={scale=0.8, transform shape}},
				y filter/.code={\pgfmathparse{1000*\pgfmathresult}},
				xmin = -1,
				xmax = 41,
				ymin = -0.02,
				ymax = 0.34,
				]
				\addplot[line width = 1.5pt, color=darkgreen] table {data_fig_5b_homhbfem.txt};
				\addplot[line width = 1.5pt, color=darkred,dashed] table {data_fig_5b_cst.txt};
			\end{axis}
		\end{tikzpicture}
		\caption{$f_{\mathrm{f}} = \SI{100}{\hertz}$.}
		\label{fig:tm_energy_100Hz}
	\end{subfigure}
	
	\vspace{0.5cm}
	
	% Second row of figures
	\begin{subfigure}{0.23\textwidth}
		\centering
		\begin{tikzpicture}[scale=0.51]
			\begin{axis}[
				xlabel = {$t\,(\si{\milli\second})$},
				xlabel style={ yshift = 5pt, font=\large},
				ylabel= {$W_{\mathrm{mag}}\,(\si{\milli\joule})$},
				ylabel style={ yshift = -10pt, font=\large},
				yticklabel style={/pgf/number format/fixed},
				grid = both,
				minor grid style = {gray!15},
				legend style={at={(0.05,0.1)},anchor = west,nodes={scale=0.8, transform shape}},
				y filter/.code={\pgfmathparse{1000*\pgfmathresult}},
				xmin = -0.2,
				xmax = 8.2,
				ymin = -0.02,
				ymax = 0.34,
				]
				\addplot[line width = 1.5pt, color=darkgreen] table {data_fig_5c_homhbfem.txt};
				\addplot[line width = 2pt, color=darkred,dashed] table {data_fig_5c_cst.txt};
			\end{axis}
		\end{tikzpicture}
		\caption{$f_{\mathrm{f}} = \SI{500}{\hertz}$.}
		\label{fig:tm_energy_500Hz}
	\end{subfigure}
	\hspace{0.1cm}
	\begin{subfigure}{0.23\textwidth}
		\centering
		\begin{tikzpicture}[scale=0.51]
			\begin{axis}[
				xlabel = {$t\,(\si{\milli\second})$},
				xlabel style={ yshift = 5pt, font=\large},
				ylabel= {$W_{\mathrm{mag}}\,(\si{\milli\joule})$},
				ylabel style={ yshift = -10pt, font=\large},
				yticklabel style={/pgf/number format/fixed},
				grid = both,
				minor grid style = {gray!15},
				legend style={at={(0.05,0.1)},anchor = west,nodes={scale=0.8, transform shape}},
				y filter/.code={\pgfmathparse{1000*\pgfmathresult}},
				xmin = -0.1,
				xmax = 4.1,
				ymin = -0.02,
				ymax = 0.34,
				]
				\addplot[line width = 1.5pt, color=darkgreen] table {data_fig_5d_homhbfem.txt};
				\addplot[line width = 2pt, color=darkred,dashed] table {data_fig_5d_cst.txt};
			\end{axis}
		\end{tikzpicture}
		\caption{$f_{\mathrm{f}} = \SI{1}{\kilo\hertz}$.}
		\label{fig:tm_energy_1000Hz}
	\end{subfigure}
	\caption{Magnetic energies in the ferromagnetic core computed with the \mbox{HomHBFEM-1} at four fundamental frequencies of the source current, compared to the transient reference results.}
	\label{fig:tm_energy__without_trafo_pos_1}
\end{figure}    
The described effect, which prevents the energy from vanishing, cannot be directly captured by the homogenization technique because this technique averages the components of flux densities parallel to the lamination over the lamination thickness. Hence, a phase shift within one lamination cannot be considered. 

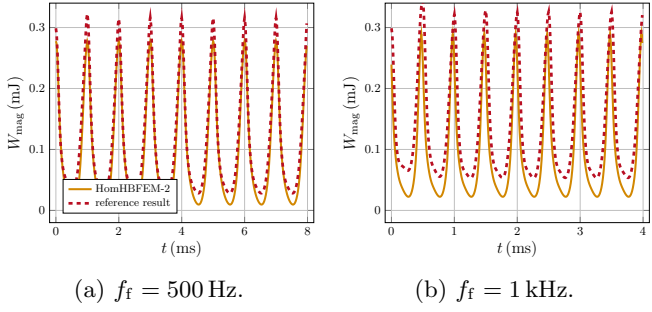
\begin{figure}[H]
	\centering
	\begin{subfigure}{0.23\textwidth}
		\centering
		\begin{tikzpicture}[scale=0.51]
			\begin{axis}[
				xlabel = {$t\,(\si{\milli\second})$},
				xlabel style={yshift = 5pt,font=\large},
				ylabel= {$W_{\mathrm{mag}}\,(\si{\milli\joule})$},
				ylabel style={ yshift = -10pt, font=\large},
				yticklabel style={/pgf/number format/fixed},
				grid = both,
				minor grid style = {gray!15},
				legend style={at={(0.05,0.12)},anchor = west,nodes={scale=0.9, transform shape}},
				y filter/.code={\pgfmathparse{1000*\pgfmathresult}},
				xmin = -0.2,
				xmax = 8.2,
				ymin = -0.02,
				ymax = 0.34,
				]
				\addplot[line width = 1.5pt, color=darkyellow] table {data_fig_6a_homhbfem.txt};
				\addlegendentry{HomHBFEM-2};
				\addplot[line width = 2pt, color=darkred,dashed] table {data_fig_6a_cst.txt};
				\addlegendentry{reference result};
			\end{axis}
		\end{tikzpicture}
		\caption{$f_{\mathrm{f}} = \SI{500}{\hertz}$.}
		\label{fig:tm_energy_500Hz_with_trafo}
	\end{subfigure}
	\hspace{0.1cm}
	\begin{subfigure}{0.23\textwidth}
		\centering
		\begin{tikzpicture}[scale=0.51]
			\begin{axis}[
				xlabel = {$t\,(\si{\milli\second})$},
				xlabel style={yshift = 5pt, font=\large},
				ylabel= {$W_{\mathrm{mag}}\,(\si{\milli\joule})$},
				ylabel style={ yshift = -10pt, font=\large},
				yticklabel style={/pgf/number format/fixed},
				grid = both,
				minor grid style = {gray!15},
				legend style={at={(0.05,0.1)},anchor = west,nodes={scale=0.8, transform shape}},
				y filter/.code={\pgfmathparse{1000*\pgfmathresult}},
				xmin = -0.1,
				xmax = 4.1,
				ymin = -0.02,
				ymax = 0.34,
				]
				\addplot[line width = 1.5pt, color=darkyellow] table {data_fig_6b_homhbfem.txt};
				%\addlegendentry{Hom. HBFEM};
				\addplot[line width = 2pt, color=darkred,dashed] table {data_fig_6b_cst.txt};
				%\addlegendentry{CST Transient};
			\end{axis}
		\end{tikzpicture}
		\caption{$f_{\mathrm{f}} = \SI{1}{\kilo\hertz}$.}
		\label{fig:tm_energy_1000Hz_with_trafo}
	\end{subfigure}
	\caption{Magnetic energies in the ferromagnetic core computed with the \mbox{HomHBFEM-2} at two fundamental frequencies of the source current, compared to the transient reference results.}
	\label{fig:tm_energy_side_by_side}
\end{figure}

In order to reach a good agreement of the magnetic energies at the higher frequencies, we need to address the issue of the phase shifts within one lamination. To that end, we need to transform the averaged flux densities parallel to the lamination, meaning the $x$- and $y$-components, $\underline{B}_{\parallel}^{\mathrm{av}}(z)$, which we obtain directly from our FE solution, back to local flux densities $B_{\parallel}(z)$. As described in~\cite{dular3DMagneticVector2003}, this can be done via
\begin{equation}
	\underline{B}_{\parallel}(z) =  \frac{\underline{B}_{\parallel}^{\mathrm{av}} (z) k d }{2 \sinh\left(k \frac{d}{2} \right)} \cosh\left( k \Tilde{z}\right),
\end{equation}
with $k = \frac{1 + \jmath}{\delta}$, and $\Tilde{z}$ being the local coordinate within one lamination, where the origin of the local coordinate system is chosen in the center of the lamination. This transformation is straightforward to implement in the linear case, which is treated in~\cite{dular3DMagneticVector2003}, but it is more difficult in the nonlinear case, because $k$ depends on the skin depth $\delta$, which in turn depends on the reluctivity and thus depends on the local flux densities themselves. Therefore, we have 
\begin{equation}
	\underline{B}_{\parallel}(z) = \frac{\underline{B}_{\parallel}^{\mathrm{av}}(z)  k( \underline{B}_{\parallel}(z) ) d}{2 \sinh\left(k (\underline{B}_{\parallel}(z)) \frac{d}{2} \right)} \cosh\left( k ( \underline{B}_{\parallel}(z)) \Tilde{z}\right). \label{eq:non_linear_transformation}
\end{equation}
Furthermore, we are dealing with multiple harmonics. Hence, Eq.~\eqref{eq:non_linear_transformation} leads to coupled nonlinear systems of equations for the different harmonics. We have two options to integrate this transformation into our method:
\begin{enumerate}
	\item \label{item:transformation_1} Perform the transformation in every iteration and evaluate $k(\underline{B}_{\parallel}(z))$ with the $\underline{B}_{\parallel}(z)$ from the previous iteration.
	\item \label{item:transformation_2} Solve the nonlinear system of equations resulting from Eq.~\eqref{eq:non_linear_transformation} once after the iteration. 
\end{enumerate}
We have found that solving the nonlinear system once after the iteration is impractical. Therefore, we will focus on the first option, which we will refer to as \mbox{HomHBFEM-2}. The results for the energies at ${f_{\mathrm{f}} = \SI{500}{Hz}}$ and ${f_{\mathrm{f}} = \SI{1}{\kilo\hertz}}$ are shown in Figs.~\ref{fig:tm_energy_500Hz_with_trafo} and~\ref{fig:tm_energy_1000Hz_with_trafo}, respectively. We can observe that including the transformation formula in our iteration did actually cause an important qualitative improvement: the magnetic energy computed with HomHBFEM does no longer go down to zero. However, further improvement is necessary.
The results can be further improved by adapting the homogenization more to the nonlinear case. One apparent remaining problem is the evaluation of the skin depth $\delta$, for which we have used the DC component of the reluctivity $\nu_0$. This is of course a potentially very rough estimation, especially at high saturation. We have found that using some sort of effective reluctivity instead of $\nu_0$ can be very beneficial. Many choices for such an effective reluctivity are available in literature, and most of them have had a positive impact on our results, in terms of reaching closer agreement with the transient reference results. A good overview of available choices is given in~\cite{paoli_time_1998}. In this paper, we focus on the following two examples.

The first approach is given in~\cite{demerdash_new_1974}. The main idea is that the effective reluctivity $\nu_{\mathrm{eff}}^{(1)}\left(\vec{r} \right)$ should fulfill 
\begin{align}
	\frac{1}{ T_{\mathrm{f}} } \int_{0}^{ T_{\mathrm{f}} } \frac{1}{2}\nu_{\mathrm{eff}}^{(1)}\left(\vec{r} \right) & | \vec{B}\left( \vec{r},t\right) |^2 \d t  \notag\\  &=\underbrace{\frac{1}{T_{\mathrm{f}} }\int_{0}^{T} \int_{0}^{\vec{B} \left(\vec{r},t \right)} H(B) \d B \,\d t}_{\overline{w}_{\mathrm{mag}}\left( \vec{r} \right)},\label{eq:eff_reluctivity_mean}
\end{align}
where $\overline{w}_{\mathrm{mag}}\left( \vec{r} \right)$ is the time-averaged magnetic energy density. This condition leads to 
\begin{equation}
	\nu_{\mathrm{eff}}^{(1)} \left( \vec{r} \right) = \frac{2 \overline{w}_{\mathrm{mag}}\left( \vec{r} \right)}{ \frac{1}{T_{\mathrm{f}}} \int_{0}^{T_{\mathrm{f}} }| \vec{B}\left( \vec{r},t\right) |^2 \d t  }.
\end{equation}
The second approach is given in~\cite{hedia_sinusoidal_1995}, where it is proposed to use
\begin{equation}
	\nu_{\mathrm{eff}}^{(2)}\left( \vec{r} \right)  = \frac{2 \int_{0}^{B_{\mathrm{max}}\left( \vec{r} \right) } H(B) dB}{{B_{\mathrm{max}}^2 \left( \vec{r} \right)}}, \label{eq:eff_reluctivity}
\end{equation}
with $B_{\mathrm{max}}\left( \vec{r} \right) = \underset{t}{\max} |\vec{B} \left(\vec{r}, t\right)| $. Both are intuitive choices based on the magnetic energy and are straightforward to integrate into our iteration scheme. Motivated by the observation that using $\nu_{\mathrm{eff}}^{(1)}$ leads to an overestimation of the energy minima and $\nu_{\mathrm{eff}}^{(2)}$ to a slight underestimation, we use the average of both approaches, i.e., 
\begin{equation}
	\nu_{\mathrm{eff}}\left( \vec{r} \right) =  \frac{\nu_{\mathrm{eff}}^{(1)}\left( \vec{r} \right)   +  \nu_{\mathrm{eff}}^{(2)} \left( \vec{r} \right) }{2}
\end{equation}
as the effective reluctivity. Note that a similar reasoning can be found in~\cite{labridis_calculation_1989}, where the average of two estimates for an effective permeability based on the magnetic co-energy is used. Incorporating the effective reluctivity into the \mbox{HomHBFEM-2} yields the second and final modification of the method which we will refer to as \mbox{HomHBFEM-3}.

The results for the magnetic energy are shown in Fig.~\ref{fig:energies_with_trafo_with_reluctivity_best}. Clearly, incorporating the effective reluctivity leads to a significant improvement. If we take a closer look at the minima of the energy, we observe a small oscillation in the HomHBFEM results, but this is mitigated if we include more harmonics in the HomHBFEM. Figure~\ref{fig:energies_final_overview} shows the results at $f_{\mathrm{f}} = \SI{1}{\kilo\hertz}$ for all three considered versions of the method compared to the transient reference results. Herein, we increased the maximum harmonic order from $m = 9$ to $m = 11$ in the \mbox{HomHBFEM-3}, to show the reduction of the oscillation in the energy minima. 
\begin{figure}[H]
	\centering
	\begin{subfigure}{0.23\textwidth}
		\centering
		\begin{tikzpicture}[scale=0.51]
			\begin{axis}[
				xlabel = {$t\,(\si{\milli\second})$},
				xlabel style={yshift = 5pt, font=\large},
				ylabel= {$W_{\mathrm{mag}}\,(\si{\milli\joule})$},
				ylabel style={ yshift = -10pt, font=\large},
				yticklabel style={/pgf/number format/fixed},
				grid = both,
				minor grid style = {gray!15},
				legend style={at={(0.05,0.12)},anchor = west,nodes={scale=0.9, transform shape}},
				y filter/.code={\pgfmathparse{1000*\pgfmathresult}},
				xmin = -0.2,
				xmax = 8.2,
				ymin = -0.02,
				ymax = 0.34,
				]
				\addplot[line width = 1.5pt, color=darkblue_two] table {data_fig_7a_homhbfem.txt};
				\addlegendentry{HomHBFEM-3};
				\addplot[line width = 2pt, color=darkred,dashed] table {data_fig_7a_cst.txt};
				\addlegendentry{reference result};
			\end{axis}
		\end{tikzpicture}
		\caption{$f_{\mathrm{f}} = \SI{500}{\hertz}$.}
		\label{fig:tm_energy_500Hz_with_trafo_with_effective_reluctivity}
	\end{subfigure}
	\hspace{0.1cm}
	\begin{subfigure}{0.23\textwidth}
		\centering
		\begin{tikzpicture}[scale=0.51]
			\begin{axis}[
				xlabel = {$t\,(\si{\milli\second})$},
				xlabel style={yshift = 5pt, font=\large},
				ylabel= {$W_{\mathrm{mag}}\,(\si{\milli\joule})$},
				ylabel style={ yshift = -10pt, font=\large},
				yticklabel style={/pgf/number format/fixed},
				grid = both,
				minor grid style = {gray!15},
				legend style={at={(0.05,0.1)},anchor = west,nodes={scale=0.8, transform shape}},
				y filter/.code={\pgfmathparse{1000*\pgfmathresult}},
				xmin = -0.1,
				xmax = 4.1,
				ymin = -0.02,
				ymax = 0.34,
				]
				\addplot[line width = 1.5pt, color=darkblue_two] table {data_fig_7b_homhbfem.txt};
				%\addlegendentry{Hom. HBFEM};
				\addplot[line width = 2pt, color=darkred,dashed] table {data_fig_7b_cst.txt};
				%\addlegendentry{CST Transient};
			\end{axis}
		\end{tikzpicture}
		\caption{$f_{\mathrm{f}} = \SI{1}{\kilo\hertz}$.}
		\label{fig:tm_energy_1000Hz_with_trafo_with_eff_reluctivity_best}
	\end{subfigure}
	\caption{Magnetic energies in the ferromagnetic core computed with the \mbox{HomHBFEM-3} at two fundamental frequencies of the source current, compared to the transient reference results.}
	\label{fig:energies_with_trafo_with_reluctivity_best}
\end{figure}
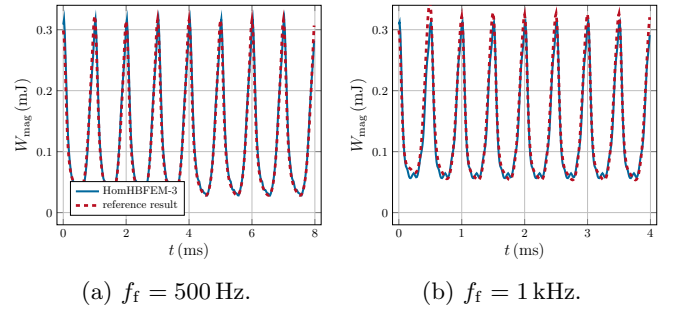
\begin{figure}[H]
	\centering
	\begin{tikzpicture}[scale=0.75]
		\begin{axis}[
			xlabel = {$t\,(\si{\milli\second})$},
			xlabel style={yshift = 5pt, font=\large},
			ylabel= {$W_{\mathrm{mag}}\,(\si{\milli\joule})$},
			ylabel style={ yshift = -10pt, font=\large},
			yticklabel style={/pgf/number format/fixed},
			grid = both,
			minor grid style = {gray!15},
			legend style={at={(0.05,0.17)},anchor = west,nodes={scale=0.9, transform shape}},
			y filter/.code={\pgfmathparse{1000*\pgfmathresult}},
			xmin = -0.1,
			xmax = 4.1,
			ymin = -0.02,
			ymax = 0.34,
			]	
			\addplot[line width = 1.5pt, color=darkgreen] table {data_fig_8_homhbfem1.txt};
			\addlegendentry{HomHBFEM-1};
			\addplot[line width = 1.5pt, color=darkyellow] table {data_fig_8_homhbfem2.txt};
			\addlegendentry{HomHBFEM-2};
			\addplot[line width = 1.5pt, color=darkblue_two] table {data_fig_8_homhbfem3.txt};
			\addlegendentry{HomHBFEM-3};
			\addplot[line width = 2pt, color=darkred,dashed] table {data_fig_8_cst.txt};
			\addlegendentry{reference result};
		\end{axis}
	\end{tikzpicture}
	\caption{Magnetic energies in the ferromagnetic core computed with all three variants of the \mbox{HomHBFEM} at $\ff = \SI{1}{\kilo\hertz}$, compared to the transient reference results.}
	\label{fig:energies_final_overview}
\end{figure}
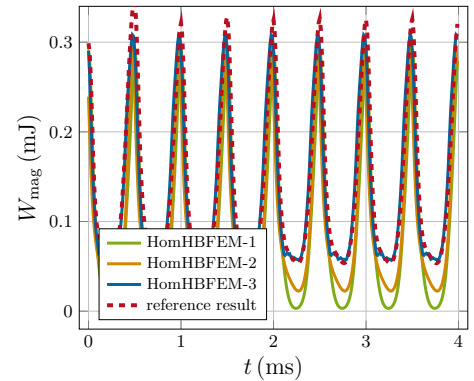
Note that the increased difference in the second energy maximum arises from transient behavior in the reference solution that is induced by its initial conditions, i.e., it is not a physically meaningful effect. 
\subsubsection{Magnetic Flux Densities}
Next, we analyze the magnetic flux densities inside the ferromagnetic core. We apply again the source current given in Eq.~\ref{eq:excitation_current} at frequencies ${f_{\mathrm{f}} \in \{50, 100, 500, 1000\}\,\si{Hz}}$. We probe the solution at different locations in the ferromagnetic core, which are specified in Table~\ref{tab:locations} in the appendix. Figure~\ref{fig:tm_flux_densities} compares the magnetic flux densities computed with the \mbox{HomHBFEM-1} to the transient reference results. Clearly and as expected based on the energy results, we see significant deviations in the magnetic flux densities for $f_{\mathrm{f}} = \SI{500}{\hertz}$ and $f_{\mathrm{f}} = \SI{1}{\kilo\hertz}$.
\begin{figure}[b]
	\centering
	% First row
	\begin{subfigure}[t]{0.23\textwidth}
		\centering
		\begin{tikzpicture}[scale=0.51]
			\begin{axis}[
				xlabel = {$t\,(\si{\milli\second})$},
				xlabel style={ yshift = 5pt,font=\large},
				ylabel= {$B\,(\si{\tesla})$},
				ylabel style={ yshift = -10pt,font=\large},
				grid = both,
				minor grid style = {gray!15},
				legend style={at={(0.05,0.12)},anchor = west,nodes={scale=0.9, transform shape}},
				x coord trafo/.code={\pgfmathparse{#1*1000}\pgfmathresult},
				xmax = 20.5e-3,
				xmin = -0.5e-3,				
				]
				\addplot[line width = 1.5pt, color=darkgreen] table {data_fig_9a_homhbfem.txt};
				\addlegendentry{HomHBFEM-1};
				\addplot[line width = 2pt, color=darkred,dashed] table {data_fig_9a_cst.txt};
				\addlegendentry{reference result};
			\end{axis}
		\end{tikzpicture}
		\caption{$f_{\mathrm{f}} = \SI{50}{\hertz}$.}
		\label{fig:FluxDensity50Hz}
	\end{subfigure}
	\vspace{1em}
	\hspace{0.1cm}
	\begin{subfigure}[t]{0.23\textwidth}
		\centering
		\begin{tikzpicture}[scale=0.51]
			\begin{axis}[
				xlabel = {$t\,(\si{\milli\second})$},
				xlabel style={ yshift = 5pt,font=\large},
				ylabel= {$B\,(\si{\tesla})$},
				ylabel style={ yshift = -10pt,font=\large},
				grid = both,
				minor grid style = {gray!15},
				legend style={at={(0.05,0.1)},anchor = west,nodes={scale=0.8, transform shape}},
				x coord trafo/.code={\pgfmathparse{#1*1000}\pgfmathresult},
				xmin = -0.25e-3,
				xmax = 10.25e-3,
				]
				\addplot[line width = 1.5pt, color=darkgreen] table {data_fig_9b_homhbfem.txt};
				\addplot[line width = 2pt, color=darkred,dashed] table {data_fig_9b_cst.txt};
			\end{axis}
		\end{tikzpicture}
		\caption{$f_{\mathrm{f}} = \SI{100}{\hertz}$.}
		\label{fig:FluxDensity100Hz}
	\end{subfigure}
	%\vspace{10em}
	% Second row
	\begin{subfigure}[t]{0.23\textwidth}
		\centering
		\begin{tikzpicture}[scale=0.51]
			\begin{axis}[
				xlabel = {$t\,(\si{\milli\second})$},
				xlabel style={ yshift = 5pt,font=\large},
				ylabel= {$B\,(\si{\tesla})$},
				ylabel style={ yshift = -10pt,font=\large},
				grid = both,
				minor grid style = {gray!15},
				legend style={at={(0.05,0.1)},anchor = west,nodes={scale=0.8, transform shape}},
				x coord trafo/.code={\pgfmathparse{#1*1000}\pgfmathresult},
				xmin = -0.05e-03,
				xmax = 2.05e-03,
				]
				\addplot[line width = 1.5pt, color=darkgreen] table {data_fig_9c_homhbfem.txt};
				\addplot[line width = 2pt, color=darkred,dashed] table {data_fig_9c_cst.txt};
			\end{axis}
		\end{tikzpicture}
		\caption{$f_{\mathrm{f}} = \SI{500}{\hertz}$.}
		\label{fig:FluxDensity500Hz}
	\end{subfigure}
	\hspace{0.1cm}
	\begin{subfigure}[t]{0.23\textwidth}
		\centering
		\begin{tikzpicture}[scale=0.51]
			\begin{axis}[
				xlabel = {$t\,(\si{\milli\second})$},
				xlabel style={ yshift = 5pt,font=\large},
				ylabel= {$B\,(\si{\tesla})$},
				ylabel style={ yshift = -10pt,font=\large},
				grid = both,
				minor grid style = {gray!15},
				legend style={at={(0.05,0.1)},anchor = west,nodes={scale=0.8, transform shape}},
				x coord trafo/.code={\pgfmathparse{#1*1000}\pgfmathresult},
				xmin = -0.025e-03,
				xmax  = 1.025e-03,
				]
				\addplot[line width = 1.5pt, color=darkgreen] table {data_fig_9d_homhbfem.txt};
				\addplot[line width = 2pt, color=darkred,dashed] table {data_fig_9d_cst.txt};
			\end{axis}
		\end{tikzpicture}
		\caption{$f_{\mathrm{f}} = \SI{1}{\kilo\hertz}$.}
		\label{fig:FluxDensity1000Hz}
	\end{subfigure}
	\caption{Magnetic flux densities in location 1 in the ferromagnetic core computed with the \mbox{HomHBFEM-1} at four fundamental frequencies of the source current, compared to the transient reference results.}
	\label{fig:tm_flux_densities}
\end{figure}
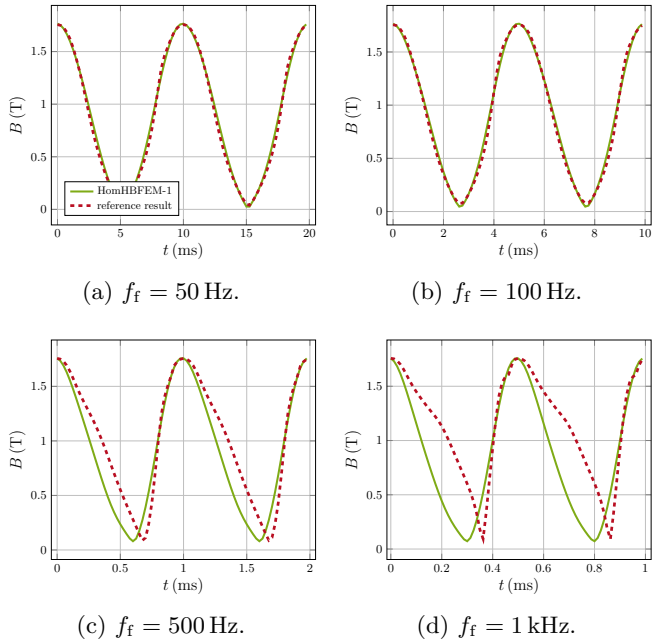
\begin{figure}[h]
	\centering
	% First row
	\begin{subfigure}[t]{0.23\textwidth}
		\centering
		\begin{tikzpicture}[scale=0.51]
			\begin{axis}[
				xlabel = {$t\,(\si{\milli\second})$},
				xlabel style={ yshift = 5pt,font=\large},
				ylabel= {$B\,(\si{\tesla})$},
				ylabel style={ yshift = -10pt,font=\large},
				grid = both,
				minor grid style = {gray!15},
				legend style={at={(0.05,0.15)},anchor = west,nodes={scale=0.9, transform shape}},
				x coord trafo/.code={\pgfmathparse{#1*1000}\pgfmathresult},
				xmin = -0.025e-03,
				xmax  = 1.025e-03,
				]
				\addplot[line width = 1.5pt, color=darkgreen] table {data_fig_10a_homhbfem1.txt};
				\addlegendentry{HomHBFEM-1};
				\addplot[line width = 1.5pt, color=darkyellow] table {data_fig_10a_homhbfem2.txt};
				\addlegendentry{HomHBFEM-2};
				\addplot[line width = 2pt, color=darkred,dashed] table {data_fig_10a_cst.txt};
				\addlegendentry{reference result};
			\end{axis}
		\end{tikzpicture}
		\caption{Location 2.}
		\label{fig:tm_flux_density_1000Hz_with_trafo_1}
	\end{subfigure}
	\vspace{1em}
	\hspace{0.1cm}
	\begin{subfigure}[t]{0.23\textwidth}
		\centering
		\begin{tikzpicture}[scale=0.51]
			\begin{axis}[
				xlabel = {$t\,(\si{\milli\second})$},
				xlabel style={ yshift = 5pt,font=\large},
				ylabel= {$B\,(\si{\tesla})$},
				ylabel style={ yshift = -10pt,font=\large},
				grid = both,
				minor grid style = {gray!15},
				legend style={at={(0.05,0.1)},anchor = west,nodes={scale=0.8, transform shape}},
				x coord trafo/.code={\pgfmathparse{#1*1000}\pgfmathresult},
				xmin = -0.025e-03,
				xmax  = 1.025e-03,
				]
				\addplot[line width = 1.5pt, color=darkgreen] table {data_fig_10b_homhbfem1.txt};
				%\addlegendentry{HomHBFEM-1};
				\addplot[line width = 1.5pt, color=darkyellow] table {data_fig_10b_homhbfem2.txt};
				%\addlegendentry{HomHBFEM-2};
				\addplot[line width = 2pt, color=darkred,dashed] table {data_fig_10b_cst.txt};
				%\addlegendentry{CST Transient};
			\end{axis}
		\end{tikzpicture}
		\caption{Location 3.}
		\label{fig:tm_flux_density_1000Hz_with_trafo_2}
	\end{subfigure}
	% Second row
	\begin{subfigure}[t]{0.23\textwidth}
		\centering
		\begin{tikzpicture}[scale=0.51]
			\begin{axis}[
				xlabel = {$t\,(\si{\milli\second})$},
				xlabel style={ yshift = 5pt,font=\large},
				ylabel= {$B\,(\si{\tesla})$},
				ylabel style={ yshift = -10pt,font=\large},
				grid = both,
				minor grid style = {gray!15},
				legend style={at={(0.05,0.1)},anchor = west,nodes={scale=0.8, transform shape}},
				x coord trafo/.code={\pgfmathparse{#1*1000}\pgfmathresult},
				xmin = -0.025e-03,
				xmax  = 1.025e-03,
				]
				\addplot[line width = 1.5pt, color=darkgreen] table {data_fig_10c_homhbfem1.txt};
				%\addlegendentry{Hom. HBFEM};
				\addplot[line width = 1.5pt, color=darkyellow] table {data_fig_10c_homhbfem2.txt};
				%\addlegendentry{Hom. HBFEM, w. trafo};
				\addplot[line width = 2pt, color=darkred,dashed] table {data_fig_10c_cst.txt};
				%\addlegendentry{CST Transient};
			\end{axis}
		\end{tikzpicture}
		\caption{Location 4.}
		\label{fig:tm_flux_density_1000Hz_with_trafo_3}
	\end{subfigure}
	%\hfill
	\hspace{0.1cm}
	\begin{subfigure}[t]{0.23\textwidth}
		\centering
		\begin{tikzpicture}[scale=0.51]
			\begin{axis}[
				xlabel = {$t\,(\si{\milli\second})$},
				xlabel style={ yshift = 5pt,font=\large},
				ylabel= {$B\,(\si{\tesla})$},
				ylabel style={ yshift = -10pt,font=\large},
				grid = both,
				minor grid style = {gray!15},
				legend style={at={(0.05,0.1)},anchor = west,nodes={scale=0.8, transform shape}},
				x coord trafo/.code={\pgfmathparse{#1*1000}\pgfmathresult},
				xmin = -0.025e-03,
				xmax  = 1.025e-03,
				]
				\addplot[line width = 1.5pt, color=darkgreen] table {data_fig_10d_homhbfem1.txt};
				%\addlegendentry{Hom. HBFEM};
				\addplot[line width = 1.5pt, color=darkyellow] table {data_fig_10d_homhbfem2.txt};
				% \addlegendentry{Hom. HBFEM, w. trafo};
				\addplot[line width = 2pt, color=darkred,dashed] table {data_fig_10d_cst.txt};
				%\addlegendentry{CST Transient};
			\end{axis}
		\end{tikzpicture}
		\caption{Location 5.}
		\label{fig:tm_flux_density_1000Hz_with_trafo_4}
	\end{subfigure}
	\caption{Magnetic flux densities in locations \mbox{2-5} in the ferromagnetic core computed with the \mbox{HomHBFEM-1} and the \mbox{HomHBFEM-2} at ${\ff = \SI{1}{\kilo\hertz}}$, compared to the transient reference results.}
	\label{fig:tm_flux_density_1000Hz_at_different_points}
\end{figure}
The benefit of including the transformation formula in the iteration, as introduced during the investigation of the magnetic energies, becomes apparent in Fig.~\ref{fig:tm_flux_density_1000Hz_at_different_points}, which shows the magnetic flux densities at four different locations inside the ferromagnetic core for the \mbox{HomHBFEM-1} and the \mbox{HomHBFEM-2} compared to the transient reference results. At this point, no effective reluctivity is included. Clearly, the transformation formula leads to much better agreement.

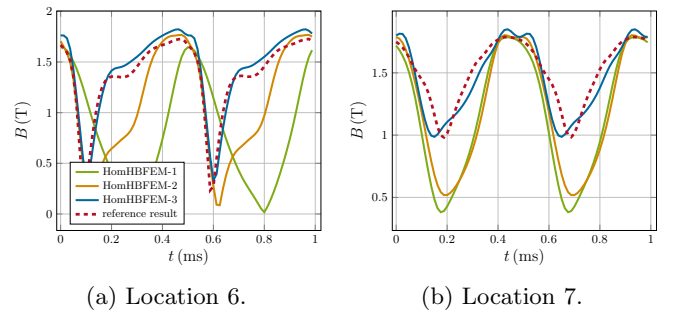
\begin{figure}[b]
	\centering
	% First row
	\begin{subfigure}[t]{0.23\textwidth}
		\centering
		\begin{tikzpicture}[scale=0.51]
			\begin{axis}[
				xlabel = {$t\,(\si{\milli\second})$},
				xlabel style={ yshift = 5pt,font=\large},
				ylabel= {$B\,(\si{\tesla})$},
				ylabel style={ yshift = -10pt,font=\large},
				grid = both,
				minor grid style = {gray!15},
				legend style={at={(0.05,0.17)},anchor = west,nodes={scale=0.9, transform shape}},
				x coord trafo/.code={\pgfmathparse{#1*1000}\pgfmathresult},
				xmin = -0.025e-03,
				xmax  = 1.025e-03,
				]
				\addplot[line width = 1.5pt, color=darkgreen] table {data_fig_11a_homhbfem1.txt};
				\addlegendentry{HomHBFEM-1};
				\addplot[line width = 1.5pt, color=darkyellow] table {data_fig_11a_homhbfem2.txt};
				\addlegendentry{HomHBFEM-2};
				\addplot[line width = 1.5pt, color=darkblue_two] table {data_fig_11a_homhbfem3.txt};
				\addlegendentry{HomHBFEM-3};
				\addplot[line width = 2pt, color=darkred,dashed] table {data_fig_11a_cst.txt};
				\addlegendentry{reference result};
			\end{axis}
		\end{tikzpicture}
		\caption{Location 6.}
		\label{fig:tm_flux_density_1000Hz_with_trafo_with_nu_eff_1}
	\end{subfigure}
	\hspace{0.1cm}
	\begin{subfigure}[t]{0.23\textwidth}
		\centering
		\begin{tikzpicture}[scale=0.51]
			\begin{axis}[
				xlabel = {$t\,(\si{\milli\second})$},
				xlabel style={ yshift = 5pt,font=\large},
				ylabel= {$B\,(\si{\tesla})$},
				ylabel style={ yshift = -10pt,font=\large},
				grid = both,
				minor grid style = {gray!15},
				legend style={at={(0.05,0.1)},anchor = west,nodes={scale=0.8, transform shape}},
				x coord trafo/.code={\pgfmathparse{#1*1000}\pgfmathresult},
				xmin = -0.025e-03,
				xmax  = 1.025e-03,
				]
				\addplot[line width = 1.5pt, color=darkgreen] table {data_fig_11b_homhbfem1.txt};
				\addplot[line width = 1.5pt, color=darkyellow] table {data_fig_11b_homhbfem2.txt};
				\addplot[line width = 1.5pt, color=darkblue_two] table {data_fig_11b_homhbfem3.txt};
				\addplot[line width = 2pt, color=darkred,dashed] table {data_fig_11b_cst.txt};
			\end{axis}
		\end{tikzpicture}
		\caption{Location 7.}
		\label{fig:tm_flux_density_1000Hz_with_trafo_with_nu_eff_2}
	\end{subfigure} 
	\caption{Magnetic flux densities in locations 6 and 7 in the ferromagnetic core computed with all three variants of the HomHBFEM at ${\ff = \SI{1}{\kilo\hertz}}$, compared to the transient reference results.}
	\label{fig:tm_flux_density_1000Hz_with_trafo_with_nu_eff_overview}
\end{figure}

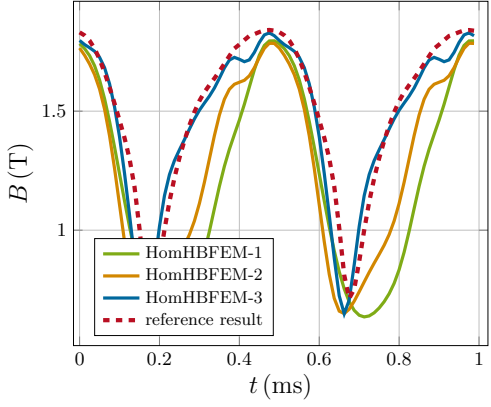
\begin{figure}[t]
	\centering
	\begin{tikzpicture}[scale=0.8]
		\begin{axis}[
			xlabel = {$t\,(\si{\milli\second})$},
			xlabel style={ yshift = 5pt,font=\large},
			ylabel= {$B\,(\si{\tesla})$},
			ylabel style={ yshift = -10pt,font=\large},
			grid = both,
			minor grid style = {gray!15},
			legend style={at={(0.05,0.17)},anchor = west,nodes={scale=0.9, transform shape}},
			x coord trafo/.code={\pgfmathparse{#1*1000}\pgfmathresult},
			xmin = -0.025e-03,
			xmax  = 1.025e-03,
			]
			\addplot[line width = 1.5pt, color=darkgreen] table {data_fig_12_homhbfem1.txt};
			\addlegendentry{HomHBFEM-1};
			\addplot[line width = 1.5pt, color=darkyellow] table {data_fig_12_homhbfem2.txt};
			\addlegendentry{HomHBFEM-2};
			\addplot[line width = 1.5pt, color=darkblue_two] table {data_fig_12_homhbfem3.txt};
			\addlegendentry{HomHBFEM-3};
			\addplot[line width = 2pt, color=darkred,dashed] table {data_fig_12_cst.txt};
			\addlegendentry{reference result};
		\end{axis}
	\end{tikzpicture}
	\caption{Magnetic flux densities in location 8 in the ferromagnetic core computed with all three variants of the HomHBFEM at ${\ff = \SI{1}{\kilo\hertz}}$, compared to the transient reference results.}
	\label{fig:tm_flux_density_1000Hz_less_drastic_changes}
\end{figure}

As for the magnetic energies, including an effective reluctivity leads to even better agreement of the HomHBFEM with the transient reference results. This is also reflected in the magnetic flux densities inside the ferromagnetic core, as shown in Fig.~\ref{fig:tm_flux_density_1000Hz_with_trafo_with_nu_eff_overview}. Note that the locations 2-7 for the comparison in Figs.~\ref{fig:tm_flux_density_1000Hz_at_different_points} and~\ref{fig:tm_flux_density_1000Hz_with_trafo_with_nu_eff_overview} have been chosen to show the improvement very clearly. In many other locations, the differences are much more subtle, as for example shown for location 8 in Fig.~\ref{fig:tm_flux_density_1000Hz_less_drastic_changes}. 

From now on, we will exclusively use the \mbox{HomHBFEM-3}, and therefore we will no longer use the numbering, i.e., we will simply refer to this final version of the method as the \mbox{HomHBFEM}. Despite the improvements made through the different versions of the method, it must be clearly stated here that a precise representation of the time signals of the flux densities at individual points within the ferromagnetic core at frequencies where the skin effect comes into play is not what this method is designed for. As shown in the plots above, the magnetic flux densities inside the laminations are roughly approximated, but there are significant differences. The method can, however, give good results on average, as indicated by the comparison of the magnetic energies. Furthermore, as we will see in the following, the HomHBFEM can give a good approximation of the eddy current losses, and most importantly for the application to the FC magnets, it can give an accurate representation of the magnetic flux density outside of the ferromagnetic material, i.e., in the aperture.
\subsubsection{Eddy Current Losses}
The eddy current losses in the ferromagnetic core consist of two contributions. First, the time-averaged losses due to the magnetic flux perpendicular to the laminations, denoted by $\LossPerpTimeAv$, and second, the time-averaged losses due to the magnetic flux parallel to the laminations, denoted by $\LossParTimeAv$. In the $\vec{A}^{*}$-formulation, the losses due to the perpendicular magnetic flux can be computed as
%\begin{align}\label{eq:LossesPerp}
%    \LossPerpTimeAv &= \int_{\OmegalsHom} \frac{1}{T_{\mathrm{f}}}\int_{0}^{T_{\mathrm{f}}} \Vec{J}\left( t \right) \cdot \Vec{E}\left( t \right) \,\d t \, \dV \notag\\ &= \int_{\OmegalsHom} \frac{1}{T_{\mathrm{f}}}\int_{0}^{T_{\mathrm{f}}}  \sigmatensor \Vec{E}\left( t \right) \cdot \Vec{E}\left( t \right) \, \d t \, \dV  \notag \\  &=   \int_{\OmegalsHom}\frac{1}{T_{\mathrm{f}}}\int_{0}^{T_{\mathrm{f}}}    \gamma \sigma_{\mathrm{c}}\left[ \left( \frac{\partial A_{x}\left( t \right) }{\partial t} \right)^{2}  +  \left( \frac{\partial A_{y}\left( t \right) }{\partial t} \right)^{2} \right]\, \d t\,\dV,
%\end{align}
\begin{align}\label{eq:LossesPerp}
	\LossPerpTimeAv &= \int_{\OmegalsHom} \frac{1}{T_{\mathrm{f}}}\int_{0}^{T_{\mathrm{f}}} \Vec{J}\left( t \right) \cdot \Vec{E}\left( t \right) \,\d t \, \dV \notag\\
	&= \int_{\OmegalsHom} \frac{1}{T_{\mathrm{f}}}\int_{0}^{T_{\mathrm{f}}}  \sigmatensor \Vec{E}\left( t \right) \cdot \Vec{E}\left( t \right) \, \d t \, \dV  \notag \\  
	&= \int_{\OmegalsHom}\frac{1}{T_{\mathrm{f}}}\int_{0}^{T_{\mathrm{f}}}    
	\gamma \sigma_{\mathrm{c}}\Bigg[ \left( \frac{\partial A_{x}\left( t \right) }{\partial t} \right)^{2}  \notag \\  
	&\qquad \qquad \qquad \qquad \quad  + \left( \frac{\partial A_{y}\left( t \right) }{\partial t} \right)^{2} \Bigg]\, \d t\,\dV,
\end{align}
where $\OmegalsHom$ denotes the homogenized lamination stack. For the losses due to the parallel magnetic flux, we have 
\begin{equation}\label{eq:LossesPar}
	\LossParTimeAv = \int_{\OmegalsHom }\frac{1}{T_{\mathrm{f}}}\int_{0}^{T_{\mathrm{f}}} \Vec{H}\left( t \right) \cdot \frac{\partial \Vec{B} \left(t \right)}{\partial t}\, \d t\, \dV.
\end{equation}
Finally, the total eddy current losses are computed as the sum of $\LossParTimeAv$ and $\LossPerpTimeAv$. Note that Eq.~\eqref{eq:LossesPar} is the physical formula for the hysteresis losses. In the context of the homogenization, however, where we do not actually consider a hysteretic material, the complex-valued reluctivity is chosen such that this term approximates the eddy current losses. Table~\ref{tab:losses_inductor} juxtaposes the eddy current losses computed with the \mbox{HomHBFEM} with the transient reference results for the losses computed with \mbox{CST Studio Suite\textsuperscript{\textregistered}} for a set of selected frequency points. 
As mentioned above, in the transient reference simulations, the mesh has to resolve the skin depth in order to achieve reliable results. This leads to up to $1.8 \cdot 10^{7}$ DoFs at ${\ff = \SI{5}{\kilo\hertz}}$. Note that for the transient reference simulation at ${\ff = \SI{10}{\kilo\hertz}}$, it was impossible with the available computing infrastructure to fully resolve the skin depth. Therefore, we used the same FE mesh as for ${\ff = \SI{5}{\kilo\hertz}}$. As a consequence, the actual losses might be a bit lower than what is indicated in Table~\ref{tab:losses_inductor}. On the other hand, to compute the model with the HomHBFEM, we used only $1.3 \cdot 10^{4}$ DoFs per harmonic component for the frequencies up to ${\ff = \SI{1}{\kilo\hertz}}$ and $2.5 \cdot 10^{4}$ DoFs per harmonic for the higher frequencies. Below ${\ff = \SI{1}{\kilo\hertz}}$, it was sufficient to consider three harmonics, at  ${\ff = \SI{2}{\kilo\hertz}}$ and $\SI{5}{\kilo\hertz}$ we increased the number of considered harmonics to five and at ${\ff = \SI{10}{\kilo\hertz}}$ to six. The agreement in the losses is sufficient, attaining a relative error between $\SI{2.9}{\percent}$ at ${\ff = \SI{1}{\kilo\hertz}}$ and ${\SI{9.9}{\percent}}$ at ${\ff = \SI{10}{\kilo\hertz}}$. 

Even for this small model, the speed-up thanks to the HomHBFEM is significant. For example, at ${\ff = \SI{1}{\kilo\hertz}}$, the simulation times are reduced from several hours to a few minutes and at ${\ff = \SI{5}{\kilo\hertz}}$ or ${\SI{10}{\kilo\hertz}}$, the simulation times are reduced from a few days to a few hours.  

\begin{table}[H]
	\centering
	\caption{Eddy current losses in the laminated inductor.}\label{tab:losses_inductor}
	\begin{tabular}{lccc} % 4 columns: first one for headers
		\toprule
		\toprule
		\multirow{2}{*}{$\ff$ (\si{\hertz})} & \multicolumn{3}{c}{Power Loss (\si{\watt})}  \\ % Row spanning columns 2 and 3
		\cmidrule{2-4} % Midrule under the spanning header
		&  HomHBFEM & Reference & Rel. Error\\ % Header row
		\midrule
		50 & $1.50 \cdot 10^{-2}$  & $1.45 \cdot 10^{-2}$& $\SI{3.4}{\percent}$  \\ % First data row
		100 & $5.68 \cdot 10^{-2}$ &  $5.47 \cdot 10^{-2}$ & $\SI{3.8}{\percent}$ \\ % Second data row
		500 & $1.19$  & $1.13$ & $\SI{5.3}{\percent}$  \\ % Third data row
		1000 & $4.00$ &  $4.12$ & $\SI{2.9}{\percent}$ \\ % Fourth data row
		2000 & $14.1$  & $14.9$ & $\SI{5.4}{\percent}$  \\ % Third data row
		5000 & $71.0$ &  $76.5$ &  $\SI{7.7}{\percent}$ \\ % Fourth data row
		10000 & $218.3$ & $242.2$ & $\SI{9.9}{\percent}$ \\ % Fourth data row
		\bottomrule
		\bottomrule
	\end{tabular}
\end{table}

\section{Verification for A C-dipole Magnet}\label{sec:c_magnet}
\subsection{Model Description}
\begin{figure}[h]
	\centering
	\begin{minipage}[t]{0.22\textwidth}
		\centering
		\includegraphics[width=0.75\textwidth]{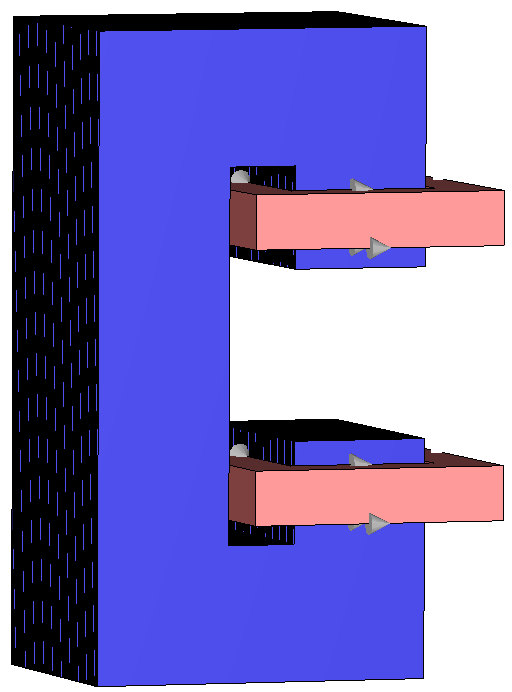}
		\caption{3D model of the C-dipole magnet.}
		\label{fig:cdipole_magnet}
	\end{minipage}
	\hspace{10pt}
	\begin{minipage}[t]{0.22\textwidth}
		\centering
		\resizebox{0.7\linewidth}{!}{\input{fig_14.tex}}
		\caption{2D cross section of the C-dipole magnet.}
		\label{fig:cdipole_magnet_cross_section}
	\end{minipage}
\end{figure}

We continue the verification studies with a dipole magnet with a C-shaped ferromagnetic yoke, as commonly used in particle accelerators, see Fig.~\ref{fig:cdipole_magnet}. This model is clearly much more realistic than the laminated inductor and could in principle be used as a FC magnet, but it is significantly smaller than the actual FC magnet to allow the computation of the transient reference results with the available computational resources. The yoke (blue) has a length of $\SI{40}{\milli\meter}$ and the transversal dimensions are summarized in Fig.~\ref{fig:cdipole_magnet_cross_section}. The lamination thickness is $d = \SI{0.5}{\milli\meter}$, the conductivity of the laminates is $\sigma = \SI{10.4}{\mega\siemens\per\meter}$, and we use the same $B$-$H$ curve as before, see Fig.~\ref{fig:BH_curve}. The total current in each of the two coils (red) is given by
\begin{equation}
	I_{\mathrm{s}}(t) = (\SI{2.5}{\kilo\ampere})\cos\left(\wf t\right).
\end{equation}
\subsection{Results}
Again, we compare the HomHBFEM results to transient reference results obtained by simulations in CST Studio Suite\textsuperscript{\textregistered}, in which the FE mesh resolves the individual laminations and, if feasible with the available computational resources, resolves the skin depth at the given frequency as well.
Here, we investigate the full frequency range of interest up to $\ff = \SI{65}{\kilo\hertz}$ in preparation for the simulation of the actual FC magnet of \mbox{PETRA IV} in the next section.
\subsubsection{Eddy Current Losses}
Figure~\ref{fig:simple_corrector_losses_comparison} shows the eddy current losses in the yoke, computed with the HomHBFEM and the transient reference simulation, over the full frequency range of interest. Having a closer look at the losses computed with the HomHBFEM, we can see that at the lower frequencies, up to roughly ${\ff = \SI{1}{\kilo\hertz}}$, they scale quadratically with the frequency. However, as we enter the kilohertz range, the dependence becomes linear. Hence, the scaling behavior of the eddy current losses computed with the HomHBFEM is in agreement with the one expected in theory~\cite{Lammeraner_1966, Stoll_1974}. The change in the scaling behavior happens because at lower frequencies, the eddy currents are restricted by the thin laminations, whereas at higher frequencies, the main limiting factor is the eddy current distribution itself, i.e., the decreasing skin depth. In the former scenario the eddy currents are referred to as being "resistance-limited", in the latter "inductance-limited"~\cite{Stoll_1974}.

\begin{figure}[H]
	\centering
	\begin{tikzpicture}[scale = 0.88]
		\begin{loglogaxis}[
			xlabel = {$\ff$\,(\si{\hertz})},
			ylabel ={$P$\,(\si{\watt})},
			xlabel style={font=\large},
			ylabel style={font=\large},
			grid = both,
			minor grid style = {gray!15},
			minor grid style = {gray!15},
			legend style={at={(0.95,0.12)},anchor = east,nodes={scale=0.9, transform shape}},
			]
			\addplot[darkblue_two,mark=*,line width=1.3pt] table[x=f(Hz), y=P_loss_av(W)] {data_fig_15_homhbfem.txt};
			\addlegendentry{HomHBFEM};
			\addplot[darkred,dashed,line width=1.3pt,mark=square, mark options={solid}] table[x=f(Hz), y=P_loss_av(W)] {data_fig_15_cst.txt};
			\addlegendentry{reference result};
		\end{loglogaxis}
	\end{tikzpicture}
	% \vspace{0.05pt}
	\caption{Eddy current losses in the C-dipole magnet's yoke as a function of frequency computed with the HomHBFEM, compared to the transient reference results.}
	\label{fig:simple_corrector_losses_comparison}
\end{figure}
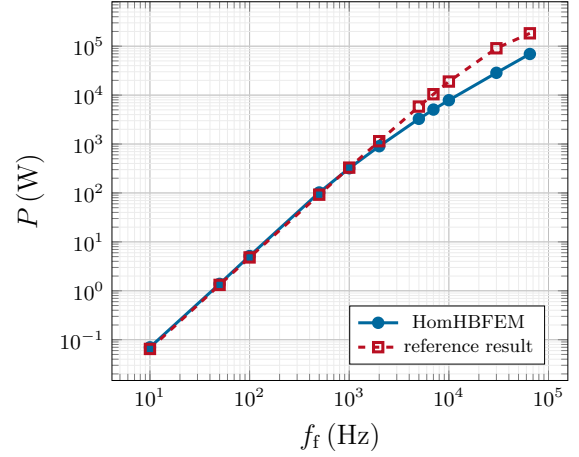

\begin{figure}[H]
	\centering
	\begin{tikzpicture}[scale=0.88]
		\begin{semilogxaxis}[
			xlabel = {\# DoFs},
			ylabel ={$P$\,(\si{\watt})},
			xlabel style={font=\large},
			ylabel style={font=\large},
			grid = both,
			xmax = 5e+7,
			xmin = 1e+5,
			minor grid style = {gray!15},
			minor grid style = {gray!15},
			] 
			\addplot[darkred,dashed, line width=1.3pt, ,mark=square, mark options={solid}] table[x=Dofs, y=P(W)] {data_fig_16.txt};
			\addplot[darkblue_two,dashed,line width=2pt] coordinates {(1e+5,320) (50000000,320)};
			\node[darkblue_two, above] at (axis cs:4e5,320) {\textbf{HomHBFEM}};
		\end{semilogxaxis}
	\end{tikzpicture}
	\caption{Eddy current losses in the C-dipole magnet's yoke at ${\ff = \SI{1}{\kilo\hertz}}$ as a function of the number of DoFs, computed with the transient reference simulation. The blue line shows the HomHBFEM results.}
	%	\caption{Eddy current losses in the C-shaped dipole magnet's yoke at ${\ff = \SI{1}{\kilo\hertz}}$ as a function of the number of DoFs, computed with CST Studio Suite\textsuperscript{\textregistered}. The blue dashed line indicates the losses computed with the HomHBFEM.}
	\label{fig:simple_corrector_losses_1kHz}
\end{figure}
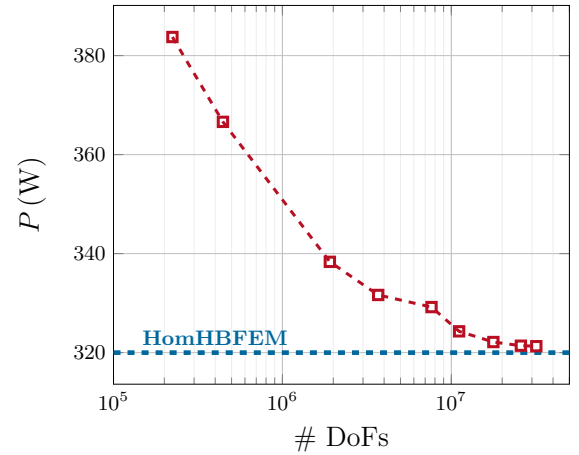

Looking at the losses computed by the transient reference simulation, we observe the same scaling behavior as for the HomHBFEM and a good agreement up to $\ff = {\SI{1}{\kilo\hertz}}$, but then, in the kilohertz range, the differences grow and the losses obtained by the transient reference simulation are significantly higher than the ones computed with the HomHBFEM.  These differences are not due to shortcomings of the HomHBFEM but are attributed to the mesh dependence of the transient reference results for the losses. The problem is that in order to obtain accurate results with the transient reference simulation, we must resolve the skin depth with the FE mesh, which is getting increasingly difficult for higher frequencies. If the skin depth is not resolved, the FEM, based on a magnetic vector potential formulation, typically overestimates the losses~\cite{meunier_2008, kaehler_2004, nikolarea_2024}.

Figure~\ref{fig:simple_corrector_losses_1kHz} shows that indeed, if we are able to sufficiently refine the mesh in the transient reference simulation, the eddy current losses converge against the losses that we have obtained with the HomHBFEM with a much coarser mesh. We have shown this in Fig.~\ref{fig:simple_corrector_losses_1kHz} for $f = \SI{1}{\kilo\hertz}$ since it is the highest frequency at which we could reach convergence of the eddy current losses in this model with CST Studio Suite\textsuperscript{\textregistered} with the available computing resources. For the HomHBFEM simulation, we included three harmonics, i.e., up to order $m = 5$, and used $1.5 \cdot 10^5$ DoFs per harmonic. On the other hand, the transient reference simulation for the right-most data point in Fig.~\ref{fig:simple_corrector_losses_1kHz} uses $3.2 \cdot 10^7$ DoFs.

For frequencies significantly above $\SI{1}{\kilo\hertz}$, where we see the large differences in Fig.~\ref{fig:simple_corrector_losses_comparison}, it is essentially impossible to reach convergence of the losses with a brute-force approach in CST Studio Suite\textsuperscript{\textregistered} with the available computing resources. This is demonstrated in Fig.~\ref{fig:simple_corrector_losses_10kHz}, which shows the eddy current losses at ${\ff = \SI{10}{\kilo\hertz}}$ over the number of DoFs in the transient reference simulation together with the corresponding simulation times on a powerful computer with $\SI{192}{\giga\byte}$ of RAM and an Intel Xeon X5690 processor. We can see that even for 16 days of simulation time, there is no convergence of the eddy current losses. By contrast, the HomHBFEM method converges with much less DoFs, which is shown in Table~\ref{tab:eddy_current_losses_HomHBFEM}. Note that the DoFs listed in Table~\ref{tab:eddy_current_losses_HomHBFEM} are only those inside of the ferromagnetic yoke, to show the refinement of the mesh more clearly. The total number of DoFs for the finest mesh are $2.53 \cdot 10^5$. Nevertheless, the HomHBFEM allows us to simulate the model in a few hours on an ordinary laptop with $\SI{16}{\giga\byte}$ of RAM and an Intel Core i7 processor. 
\begin{figure}[H]
	\centering
	\begin{tikzpicture}[scale=0.8, transform shape]
		\begin{semilogxaxis}[
			xlabel = {\# DoFs},
			x label style={at={(axis description cs:0.5,0.075)},anchor=north},
			ylabel ={$P$\,(\si{\kilo\watt})},
			xlabel style={font=\large},
			ylabel style={font=\large},
			grid = both,
			minor grid style = {gray!15},
			minor grid style = {gray!15},
			y filter/.code={\pgfmathparse{0.001*\pgfmathresult}},
			]
			\addplot[darkred, dashed, line width=1.3pt, mark=square, mark options={solid}] table[x=Dofs, y=P(W)] {data_fig_17.txt};
		\end{semilogxaxis}
		
		% Additional plot elements for simulation time annotations
		\draw[->] (0,-1.1+0.25) -- (7,-1.1+0.25);
		\node[] at (3.5,-1.8+0.25) {Simulation time};
		
		\draw[] (0.5,-1.05+0.25) -- (0.5,-1.15+0.25);
		\node[]at (0.5,-1.4+0.25) {\SI{7}{\minute}};
		
		\draw[] (2.95,-1.05+0.25) -- (2.95,-1.15+0.25);
		\node[]at (2.95,-1.4+0.25) {\SI{4}{\hour}};
		
		\draw[] (6.3,-1.05+0.25) -- (6.3,-1.15+0.25);
		\node[]at (6.3,-1.4+0.25) {\SI{16}{\day}};
	\end{tikzpicture}
	\caption{Eddy current losses in the C-dipole magnet's yoke at ${\ff = \SI{10}{\kilo\hertz}}$ as a function of the number of DoFs, computed with the transient reference simulation. The simulation times are indicated below.}
	\label{fig:simple_corrector_losses_10kHz}
\end{figure}
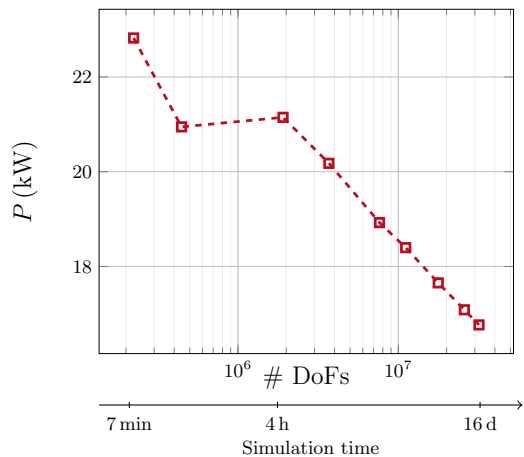
\begin{table}[H]
	\centering
	\caption{Eddy current losses in the C-dipole magnet's yoke at $\ff= \SI{10}{\kilo\hertz}$, computed with the HomHBFEM for different numbers of DoFs in the yoke.}
	\begin{tabular}{cc}
		\toprule
		\toprule
		\# DoFs & $P\,(\si{\kilo\watt})$  \\
		\midrule 
		$2.4 \cdot 10^{4}$ & $7.38$ \\
		$2.9\cdot 10^{4}$  & $7.36$ \\
		$3.6 \cdot 10^{4}$ & $7.37$ \\
		$5.6 \cdot 10^{4}$ & $7.33$ \\
		$8.0 \cdot 10^{4}$ & $7.36$ \\
		$1.08 \cdot 10^{5}$ & $7.38$ \\
		\bottomrule
		\bottomrule
	\end{tabular}
	\label{tab:eddy_current_losses_HomHBFEM}
\end{table}

\subsubsection{Magnetic Flux Densities}
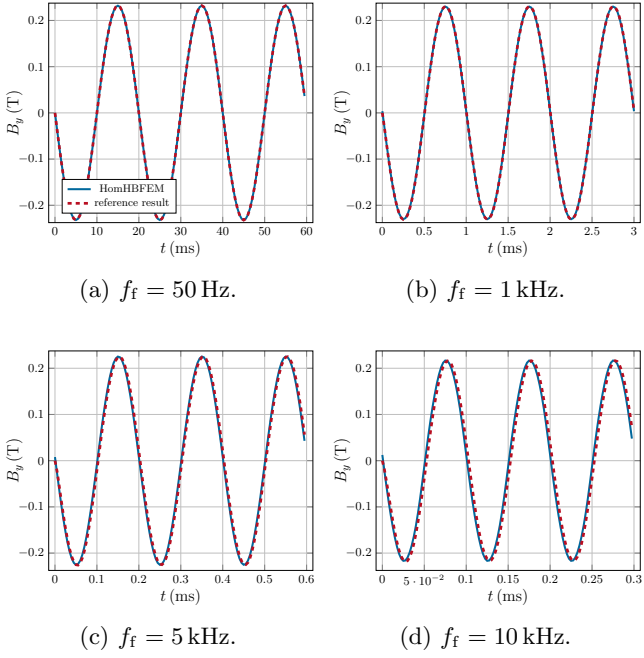
\begin{figure}[h]
	\centering
	% First row of figures
	\begin{subfigure}{0.23\textwidth}
		\centering
		\begin{tikzpicture}[scale=0.51]
			\begin{axis}[
				xlabel = {$t\,(\si{\milli\second})$},
				xlabel style={ yshift = 5pt,font=\large},
				ylabel= {$B_{y}\,(\si{\tesla})$},
				ylabel style={ yshift = -10pt,font=\large},
				grid = both,
				minor grid style = {gray!15},
				minor grid style = {gray!15},
				legend style={at={(0.05,0.12)},anchor = west,nodes={scale=0.9, transform shape}},
				x filter/.code={\pgfmathparse{1000*\pgfmathresult}},
				xmin = -1.5,
				xmax = 61.5,
				ymin = -0.2375,
				ymax = 0.2375,
				]
				\addplot[line width = 1.5pt, color=darkblue_two] table {data_fig_18a_homhbfem.txt};
				\addlegendentry{HomHBFEM};
				\addplot[line width = 2pt, color=darkred,dashed] table {data_fig_18a_cst.txt};
				\addlegendentry{reference result};
			\end{axis}
		\end{tikzpicture}
		\centering
		\caption{${\ff = \SI{50}{\hertz}}$.} \label{fig:simple_corrector_field_aperture_50Hz}
	\end{subfigure}%
	\hspace{0.1cm}
	\begin{subfigure}{0.23\textwidth}
		\centering
		\begin{tikzpicture}[scale=0.51]
			\begin{axis}[
				xlabel = {$t\,(\si{\milli\second})$},
				xlabel style={ yshift = 5pt,font=\large},
				ylabel= {$B_{y}\,(\si{\tesla})$},
				ylabel style={ yshift = -10pt,font=\large},
				grid = both,
				minor grid style = {gray!15},
				minor grid style = {gray!15},
				legend style={at={(0.05,0.1)},anchor = west,nodes={scale=0.8, transform shape}},
				x filter/.code={\pgfmathparse{1000*\pgfmathresult}},
				xmin = -0.075,
				xmax = 3.075,
				ymin = -0.2375,
				ymax = 0.2375,
				]
				\addplot[line width = 1.5pt, color=darkblue_two] table {data_fig_18b_homhbfem.txt};
				%\addlegendentry{Hom. HBFEM};
				\addplot[line width = 2pt, color=darkred,dashed] table {data_fig_18b_cst.txt};
				%\addlegendentry{CST Transient};
			\end{axis}
		\end{tikzpicture}
		\centering
		\caption{${\ff = \SI{1}{\kilo\hertz}}$.} \label{fig:simple_corrector_field_aperture_1000Hz}
	\end{subfigure}
	
	\vspace{0.5cm} % Adjust space between rows if needed
	
	% Second row of figures
	\begin{subfigure}{0.23\textwidth}
		\centering
		\begin{tikzpicture}[scale=0.51]
			\begin{axis}[
				xlabel = {$t\,(\si{\milli\second})$},
				xlabel style={ yshift = 5pt,font=\large},
				ylabel= {$B_{y}\,(\si{\tesla})$},
				ylabel style={ yshift = -10pt,font=\large},
				grid = both,
				minor grid style = {gray!15},
				minor grid style = {gray!15},
				legend style={at={(0.05,0.1)},anchor = west,nodes={scale=0.8, transform shape}},
				x filter/.code={\pgfmathparse{1000*\pgfmathresult}},
				xmin = -0.015,
				xmax = 0.615,
				ymin = -0.2375,
				ymax = 0.2375,
				]
				\addplot[line width = 1.5pt, color=darkblue_two] table {data_fig_18c_homhbfem.txt};
				%\addlegendentry{Hom. HBFEM};
				\addplot[line width = 2pt, color=darkred,dashed] table {data_fig_18c_cst.txt};
				%\addlegendentry{CST Transient};
			\end{axis}
		\end{tikzpicture}
		\centering
		\caption{${\ff = \SI{5}{\kilo\hertz}}$.} \label{fig:simple_corrector_field_aperture_5000Hz}
	\end{subfigure}%
	\hspace{0.1cm}
	\begin{subfigure}{0.23\textwidth}
		\centering
		\begin{tikzpicture}[scale=0.51]
			\begin{axis}[
				xlabel = {$t\,(\si{\milli\second})$},
				xlabel style={ yshift = 5pt,font=\large},
				ylabel= {$B_{y}\,(\si{\tesla})$},
				ylabel style={ yshift = -10pt,font=\large},
				grid = both,
				minor grid style = {gray!15},
				minor grid style = {gray!15},
				legend style={at={(0.05,0.1)},anchor = west,nodes={scale=0.8, transform shape}},
				x filter/.code={\pgfmathparse{1000*\pgfmathresult}},
				xmin = -0.0075,
				xmax = 0.3075,
				ymin = -0.2375,
				ymax = 0.2375,
				]
				\addplot[line width = 1.5pt, color=darkblue_two] table {data_fig_18d_homhbfem.txt};
				%\addlegendentry{Hom. HBFEM};
				\addplot[line width = 2pt, color=darkred,dashed] table {data_fig_18d_cst.txt};
				%\addlegendentry{CST Transient};
			\end{axis}
		\end{tikzpicture}
		\centering
		\caption{${\ff = \SI{10}{\kilo\hertz}}$.} \label{fig:simple_corrector_field_aperture_10000Hz}
	\end{subfigure}
	\caption{Magnetic flux density in the center of the C-dipole magnet's aperture computed with the HomHBFEM at four fundamental frequencies of the source current, compared to the transient reference results.}
	\label{fig:cmagnet_flux_density_aperture_center}
\end{figure}

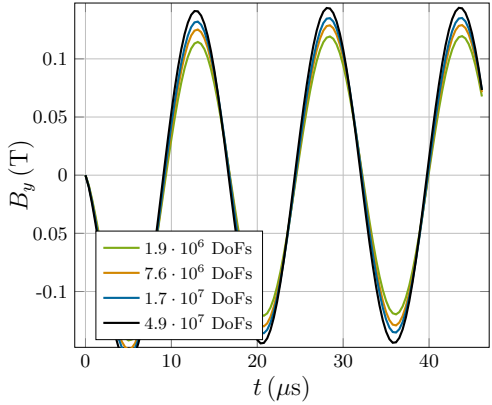
\begin{figure}[h]
	\centering
	\begin{tikzpicture}[scale=0.8]
		\begin{axis}[
			xlabel = {$t\,(\mathrm{\mu s})$},
			xlabel style={ yshift = 5pt,font=\large},
			ylabel= {$B_{y}\,(\si{\tesla})$},
			ylabel style={ yshift = -10pt,font=\large},
			grid = both,
			minor grid style = {gray!15},
			minor grid style = {gray!15},
			legend style={at={(0.05,0.18)},anchor = west,nodes={scale=0.9, transform shape}},
			x filter/.code={\pgfmathparse{1e+6*\pgfmathresult}},
			xmin = -1.25,
			xmax = 47,
			ymin = -0.148,
			ymax = 0.148,
			yticklabels={,-0.1,0.05,0,0.05,0.1,},
			]
			\addplot[line width = 1pt, color=darkgreen] table {data_fig_19_1.txt};
			\addlegendentry{$1.9 \cdot 10^6$ DoFs};
			\addplot[line width = 1pt, color=darkyellow] table {data_fig_19_2.txt};
			\addlegendentry{$7.6 \cdot 10^6$ DoFs};
			\addplot[line width = 1pt, color=darkblue_two] table {data_fig_19_3.txt};
			\addlegendentry{$1.7 \cdot 10^7$ DoFs};
			\addplot[line width = 1pt, color=black] table {data_fig_19_4.txt};
			\addlegendentry{$4.9 \cdot 10^7$ DoFs};
		\end{axis}
	\end{tikzpicture}
	\centering
	\caption{Transient reference results for the magnetic flux density in the center of the C-dipole magnet's aperture at ${\ff = \SI{65}{\kilo\hertz}}$ for four different numbers of DoFs.} \label{fig:simple_corrector_field_aperture_65000Hz}
\end{figure}

Next, we investigate the magnetic flux densities outside of the ferromagnetic material, in the center of the aperture. Figure~\ref{fig:cmagnet_flux_density_aperture_center} compares the HomHBFEM and the transient reference simulation for a source current with fundamental frequencies between $\ff = \SI{50}{\hertz}$ and $\ff = \SI{10}{\kilo\hertz}$. We observe very good agreement, with the largest relative error in the magnetic flux density's amplitude being $\SI{0.3}{\percent}$ at ${\ff = \SI{1}{\kilo\hertz}}$. If we increase the frequency significantly beyond ${\ff = \SI{10}{\kilo\hertz}}$, it again becomes virtually impossible to obtain a reliable reference solution, i.e., to reach convergence for the aperture field with a brute-force approach in CST Studio Suite\textsuperscript{\textregistered}. This is illustrated in Fig.~\ref{fig:simple_corrector_field_aperture_65000Hz}, which depicts the vertical component of the magnetic flux density in the center of the aperture at ${\ff = \SI{65}{\kilo\hertz}}$ for different FE meshes, i.e., for different numbers of DoFs. Note that the result with the finest mesh, with $4.9 \cdot 10^7$ DoFs, took 24 days of simulation time on a powerful computer with $\SI{256}{\giga\byte}$ of RAM and an Intel Xeon E5-2680 processor. 

\section{Application to Fast Corrector Magnets}\label{sec:corrector}
\subsection{Model Description}
\begin{figure}[b]
	\centering
	% First image (PNG)
	\begin{minipage}[t]{0.2\textwidth}
		\centering
		\includegraphics[width=\textwidth]{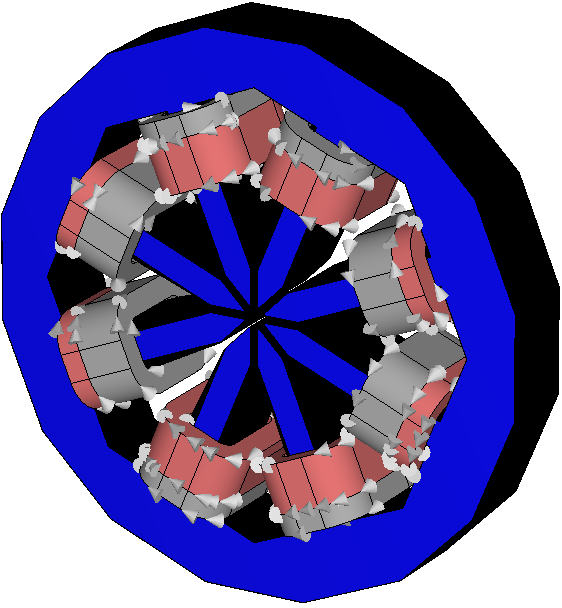}
		% \captionsetup{width=\textwidth} % Enforce consistent caption width
		\label{fig:corrector_model_3D}
	\end{minipage}
	\hspace{5pt}
	% Second image (PDF)
	\begin{minipage}[t]{0.2\textwidth}
		\centering
		\includegraphics[width=1.15\textwidth]{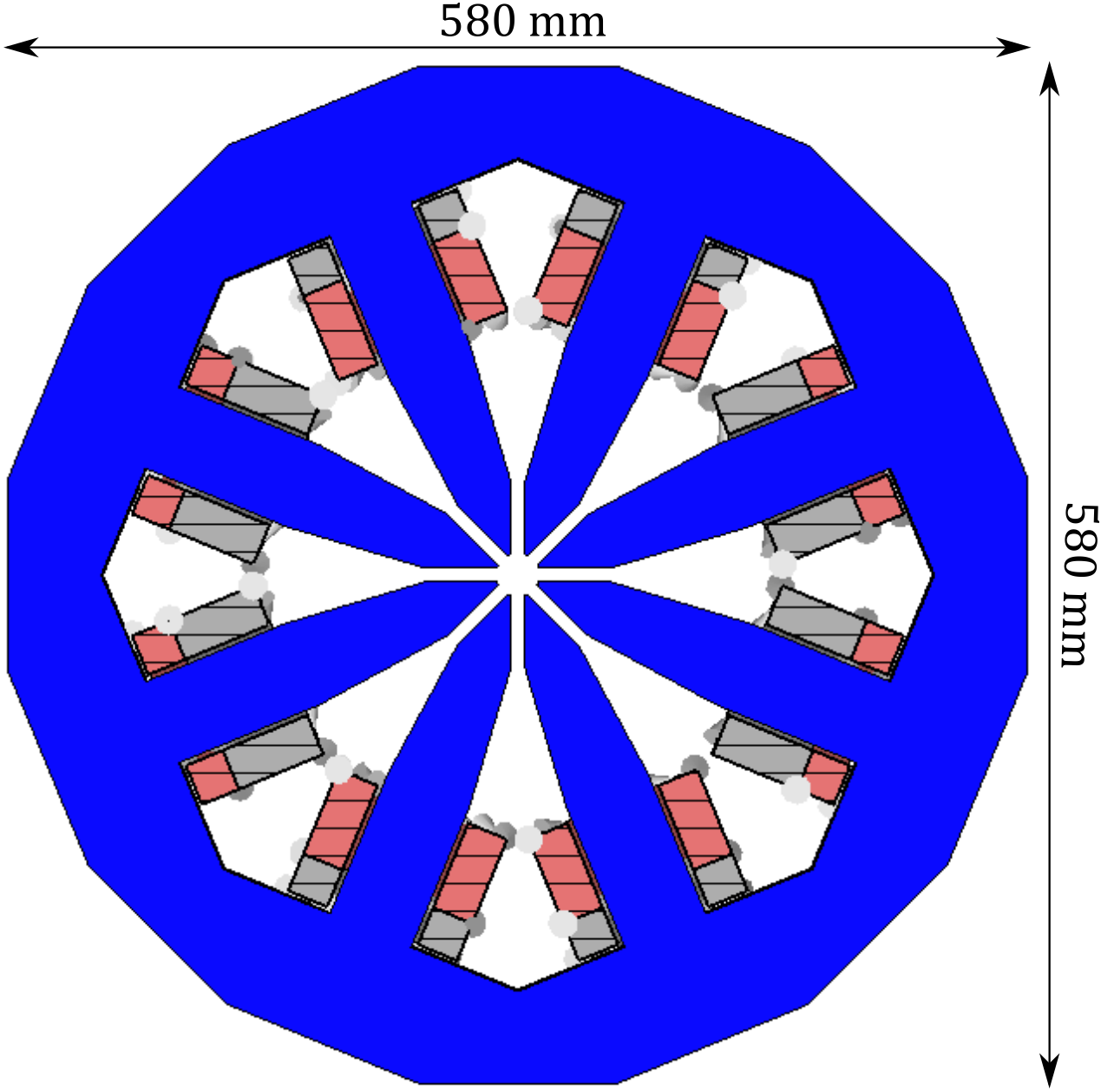}
		\label{fig:corrector_model_2D}
	\end{minipage}
	\caption{3D model of the FC magnet for PETRA IV and its cross section.}\label{fig:fc_magnet}
\end{figure}

Having derived and verified the HomHBFEM, we turn to the application to the FC magnets for PETRA IV at DESY. The 3D model and its cross section are depicted in Fig.~\ref{fig:fc_magnet}. Note that this type of octupole-like FC magnet design originates from the APS Upgrade project at the ANL~\cite{depaola_mechanical_2018}. Other synchrotron radiation sources such as SIRIUS in Brazil or HEPS in China are using or planning to use similar FC magnets with a quadrupole-like \mbox{design~\cite{giachero_2021,huang_2023}}. 

As already indicated, the magnet has an octupole-like geometry, with eight posts pointing to the aperture. However, it produces a dipole field to deflect the electrons passing through the aperture. On each of the eight posts, there is a large main coil and a smaller auxiliary coil, only one of which carries a current at any given point in time. If the red coils are powered, a vertical field is produced, if the grey coils are powered, a horizontal field is produced. In this way, the magnet can deflect the beam both in the horizontal and the vertical plane, which is important to meet the space constraints in the densely packed magnetic lattice. Furthermore, the design offers a good field quality, i.e., the particular choice of the currents allows to cancel out sextupole and decapole components in the aperture field. We consider the case that the red main coils carry $975\,\mathrm{At}$ and the red auxiliary coils carry $405\,\mathrm{At}$ and for both types of coils, the time signal of the source current is a monofrequent sinusoidal function. It should be noted here that the sinusoidal signal serves as a test signal to facilitate the analysis of the magnet's dynamic behavior.  In practice, the magnet operates as part of the FOFB, where the actual coil currents are determined by the control algorithm.

The geometrical details are given in Table~\ref{tab:corrector_geo_params}. The yoke length and diameter in Table~\ref{tab:corrector_geo_params} are slightly different from the measures of the prototype magnet that has been built recently, but we have investigated both versions of the model and the differences in terms of the investigated quantities are small. In particular, all results regarding the impact of the nonlinearity compiled below hold qualitatively for both versions of the model. We will investigate two lamination thicknesses, first $d = \SI{1}{\milli\meter}$, which is the lamination thickness of the prototype magnet and then $d = \SI{0.3}{\milli\meter}$, which is intended for series production. For a lamination thickness of $d = \SI{1}{\milli\meter}$, the stacking factor is $\gamma = \SI{98.5}{\percent}$, whereas for $\SI{0.3}{\milli\meter}$ laminations it is decreased accordingly to $\gamma = \SI{95.2}{\percent}$.

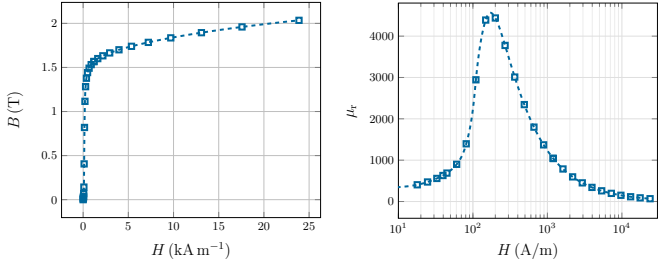
\begin{figure}[H]
	\begin{minipage}[b]{0.48\linewidth}
		\input{fig_21a.tex}
	\end{minipage}
	\hfill
	\begin{minipage}[b]{0.48\linewidth}
		\input{fig_21b.tex}
	\end{minipage}
	\caption{$B$-$H$ curve (left) and relative permeability (right) of powercore\textsuperscript{\textregistered} $1400$-$100\text{AP}$.}\label{fig:powercore_material_properties}
\end{figure}

The yoke material is powercore\textsuperscript{\textregistered} $1400$-$100\text{AP}$. The conductivity, as provided by the manufacturer thyssenkrupp, is $\SI{5.814}{\mega\siemens\per\meter}$. The \mbox{$B$-$H$ curve}, as measured at DESY, is plotted in Fig.~\ref{fig:powercore_material_properties} together with the resulting relative permeability $\mu_{\mathrm{r}} = B/ (\mu_{\mathrm{vac}} H)$, where $\mu_{\mathrm{vac}}$ is the vacuum permeability.
\begin{table}[H]
	\caption{Geometric parameters of the FC magnet.}\label{tab:corrector_geo_params}
	\centering
	\begin{tabular}{ll}
		\toprule
		\toprule
		Yoke length & $\SI{90}{\milli\meter}$   \\
		Yoke diameter &$\SI{580}{\milli\meter}$   \\
		Aperture diameter  & $\SI{25}{\milli\meter}$ \\
		Lamination thickness & $\SI{1}{\milli\meter} \, \text{or} \, \SI{0.3}{\milli\meter}$ \\
		Stacking factor &  $\SI{98.5}{\percent} \, \text{or} \, \SI{95.2}{\percent}$ \\
		\bottomrule
		\bottomrule
	\end{tabular}
\end{table}

%\begin{figure}[H]
%    \centering
%%     \resizebox{0.35\textwidth}{!}{%
	%        \input{standalone_figures/bh_powercore.tex}
	%    }
%    \caption{$B$-$H$ curve of powercore\textsuperscript{\textregistered} %$1400$-$100\text{AP}$.}\label{fig:bh_curve_powercore}
%\end{figure}
%\begin{figure}[H]
%	\centering
%	\resizebox{0.35\textwidth}{!}{%
	%		\input{standalone_figures/permeability_powercore.tex}
	%	}
%	\caption{Permeability of powercore\textsuperscript{\textregistered} $1400$-$100\text{AP}$.}\label{fig:mu_powercore}
%\end{figure}
\subsection{Results}
The most important quantities of interest are the integrated value of the magnitude of the magnetic flux density's dipole component and the phase shift between the field in the aperture and the current in the coils. The knowledge of these quantities up to the kilohertz range is crucial for modelling the FOFB of PETRA IV. In~\cite{christmann_findings}, we have computed these quantities by linear simulations to obtain an integrated transfer function for the FC magnets, which has been used at DESY for the design of the feedback control. 

\begin{figure}[H]
	\centering
	% First row of figures
	\begin{subfigure}{0.23\textwidth}
		\centering
		\begin{tikzpicture}[scale=0.51]
			\begin{axis}[
				xlabel = {$z\,(\si{\milli\meter})$},
				xlabel style={ yshift = 5pt,font=\large},
				ylabel= {$|\underline{B}_{y,1}| \,(\si{\milli\tesla})$},
				ylabel style={ yshift = -10pt,font=\large},
				%title = \textbf{Vertical Field Along the Axis, $f = \SI{10}{\hertz}$},
				grid = both,
				minor grid style = {gray!15},
				legend style={at={(0.99,0.88)},anchor = east,nodes={scale=0.9, transform shape}},
				x coord trafo/.code={\pgfmathparse{#1*1000}\pgfmathresult},
				y coord trafo/.code={\pgfmathparse{#1*1000}\pgfmathresult},
				xmin = -320e-03,
				xmax = 320e-03,
				]
				\addplot[line width = 1.5pt, color=darkblue_two] table {data_fig_22a_nl.txt};
				\addlegendentry{HomHBFEM};
				\addplot[line width = 2.25pt, color=darkyellow,dashed] table {data_fig_22a_lin.txt};
				\addlegendentry{linear result};
			\end{axis}
		\end{tikzpicture}
		\caption{$\ff = \SI{10}{\hertz}.$}
		\label{fig:field_lin_vs_nl_10Hz}
	\end{subfigure}%
	\hspace{0.1cm}
	\begin{subfigure}{0.23\textwidth}
		\centering
		\begin{tikzpicture}[scale=0.51]
			\begin{axis}[
				xlabel = {$z\,(\si{\milli\meter})$},
				xlabel style={ yshift = 5pt,font=\large},
				ylabel= {$|\underline{B}_{y,1}| \,(\si{\milli\tesla})$},
				ylabel style={ yshift = -10pt,font=\large},
				%title = \textbf{Vertical Field Along the Axis, $f = \SI{5}{\kilo\hertz}$},
				grid = both,
				minor grid style = {gray!15},
				legend style={at={(0.99,0.9)},anchor = east,nodes={scale=0.8, transform shape}},
				x coord trafo/.code={\pgfmathparse{#1*1000}\pgfmathresult},
				y coord trafo/.code={\pgfmathparse{#1*1000}\pgfmathresult},
				xmin = -320e-03,
				xmax = 320e-03,
				]
				\addplot[line width = 1.5pt, color=darkblue_two] table {data_fig_22b_nl.txt};
				%\addlegendentry{Nonlinear};
				\addplot[line width = 2.25pt, color=darkyellow,dashed] table {data_fig_22b_lin.txt};
				%\addlegendentry{Linear};
			\end{axis}
		\end{tikzpicture}
		\caption{$\ff = \SI{5}{\kilo\hertz}.$}
		\label{fig:field_lin_vs_nl_5000Hz}
	\end{subfigure}
	
	\vspace{0.5cm} % Adjust space between rows if needed
	
	% Second row of figures
	\begin{subfigure}{0.23\textwidth}
		\centering
		\begin{tikzpicture}[scale=0.51]
			\begin{axis}[
				xlabel = {$z\,(\si{\milli\meter})$},
				xlabel style={ yshift = 5pt,font=\large},
				ylabel= {$|\underline{B}_{y,1}| \,(\si{\milli\tesla})$},
				ylabel style={ yshift = -10pt,font=\large},
				% title = \textbf{Vertical Field Along the Axis, $f = \SI{10}{\kilo\hertz}$},
				grid = both,
				minor grid style = {gray!15},
				legend style={at={(0.99,0.9)},anchor = east,nodes={scale=0.8, transform shape}},
				x coord trafo/.code={\pgfmathparse{#1*1000}\pgfmathresult},
				y coord trafo/.code={\pgfmathparse{#1*1000}\pgfmathresult},
				xmin = -320e-03,
				xmax = 320e-03,
				]
				\addplot[line width = 1.5pt, color=darkblue_two] table {data_fig_22c_nl.txt};
				%\addlegendentry{Nonlinear};
				\addplot[line width = 2.25pt, color=darkyellow,dashed] table {data_fig_22c_lin.txt};
				%\addlegendentry{Linear};
			\end{axis}
		\end{tikzpicture}
		\caption{$\ff = \SI{10}{\kilo\hertz}.$}
		\label{fig:field_lin_vs_nl_10000Hz}
	\end{subfigure}%
	\hspace{0.1cm}
	\begin{subfigure}{0.23\textwidth}
		\centering
		\begin{tikzpicture}[scale=0.51]
			\begin{axis}[
				xlabel = {$z\,(\si{\milli\meter})$},
				xlabel style={ yshift = 5pt,font=\large},
				ylabel= {$|\underline{B}_{y,1}| \,(\si{\milli\tesla})$},
				ylabel style={ yshift = -10pt,font=\large},
				% title = \textbf{Vertical Field Along the Axis, $f = \SI{65}{\kilo\hertz}$},
				grid = both,
				minor grid style = {gray!15},
				legend style={at={(0.99,0.9)},anchor = east,nodes={scale=0.8, transform shape}},
				x coord trafo/.code={\pgfmathparse{#1*1000}\pgfmathresult},
				y coord trafo/.code={\pgfmathparse{#1*1000}\pgfmathresult},
				xmin = -320e-03,
				xmax = 320e-03,
				]
				\addplot[line width = 1.5pt, color=darkblue_two] table {data_fig_22d_nl.txt};
				%\addlegendentry{Nonlinear};
				\addplot[line width = 2.25pt, color=darkyellow, dashed] table {data_fig_22d_lin.txt};
				%\addlegendentry{Linear};
			\end{axis}
		\end{tikzpicture}
		\caption{$\ff = \SI{65}{\kilo\hertz}.$}
		\label{fig:field_lin_vs_nl_65000Hz}
	\end{subfigure}
	\caption{Magnitude of the dipole component of the magnetic flux density's first harmonic along the axis of the FC magnet computed with the HomHBFEM for $\SI{1}{\milli\meter}$ laminations at four fundamental frequencies of the source current, compared to the linear results.}
	\label{fig:vertical_field_along_axis}
\end{figure}
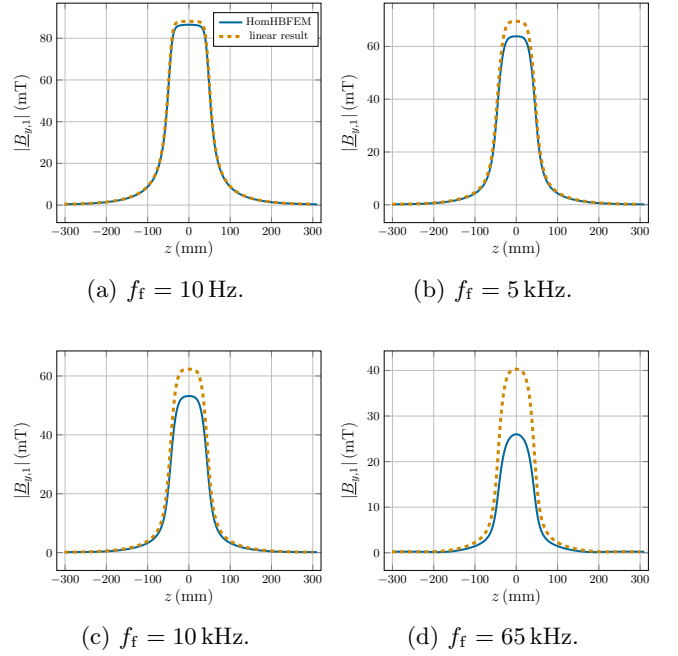

So far, conducting nonlinear simulation studies for the FC magnets at elevated frequencies had not been possible due to the entailed prohibitive computational effort, which has already become evident in the analysis of the smaller models in the previous Sections. With the HomHBFEM, we are now able to conduct nonlinear simulations in the kilohertz range, which allows us to  investigate the effect that a nonlinear $B$-$H$ curve has on the dynamic behavior of the magnet. To that end, we will compare the results of the nonlinear simulation with the HomHBFEM to the linear simulation that we had conducted previously, see e.g.~\cite{christmann_findings, christmannFiniteElementSimulation2024a}. Figure~\ref{fig:vertical_field_along_axis} shows this comparison between linear and nonlinear simulations for the longitudinal magnetic field profiles in the model with $d = \SI{1}{\milli\meter}$.
For the linear simulation, we use the homogenization technique in its basic form, as described in Sec.~\ref{sec:hom}. For the nonlinear simulation, we plot the magnitude of the dipole component of the first harmonic. Higher order harmonics are negligible in the aperture. 

For both the linear and the nonlinear simulation, we observe the typical attenuation of the magnetic flux density due to the eddy currents as the frequency is increased. Further, we see that at lower frequencies, such as ${\ff=\SI{10}{\hertz}}$, the results of the linear an the nonlinear simulation are very similar. However, at higher frequencies we see that the magnetic flux density from the nonlinear simulation is significantly lower than the one from the linear simulation. 
\begin{table}[b]
	\centering
	\caption{Integrated magnetic flux density and phase shift in the center of the magnet's aperture with $\SI{1}{\milli\meter}$ laminations.}  \label{tab:int_field_phase_shift}
	\begin{tabular}{@{}lcc|cc@{}}
		\toprule
		\toprule
		\multirow{2}{*}{$\ff$ (Hz)} & \multicolumn{2}{c|}{Linear} & \multicolumn{2}{c}{Nonlinear} \\ 
		& $B_{y, \mathrm{int}} (\si{\milli\tesla\meter})$ & $\varphi_{\mathrm{center}} (\si{\degree})$       & $B_{y, \mathrm{int}} (\si{\milli\tesla\meter})$    & $\varphi_{\mathrm{center}}  (\si{\degree})$      \\ \midrule
		$10$        & $11.7$    & $-0.1$          & $11.4$      & $0.0$          \\
		$5000$        & $7.9$    & $-13.0$          & $7.2$      & $-16.1$          \\
		$10000$        & $6.9$    & $-16.3$          & $5.8$      & $-20.2$          \\
		$65000$        & $4.2$    & $-25.4$         & $2.7$     & $-31.8$         \\
		\bottomrule
		\bottomrule
	\end{tabular}
\end{table}
This is also illustrated in Table~\ref{tab:int_field_phase_shift}, which lists the integrated values of the magnetic flux densities plotted in Fig.~\ref{fig:vertical_field_along_axis}, denoted by $B_{y, \mathrm{int}}$, as well as the respective phase shifts between the aperture field and the current in the coils, $\varphi_{\mathrm{center}}$. Note that the integration limits for the computation of $B_{y, \mathrm{int}}$ were chosen to be $z = \pm \SI{500}{\milli\meter}$. We observe that at higher frequencies, the integrated field in the nonlinear case is significantly smaller and the phase shift is greater than in the linear case.

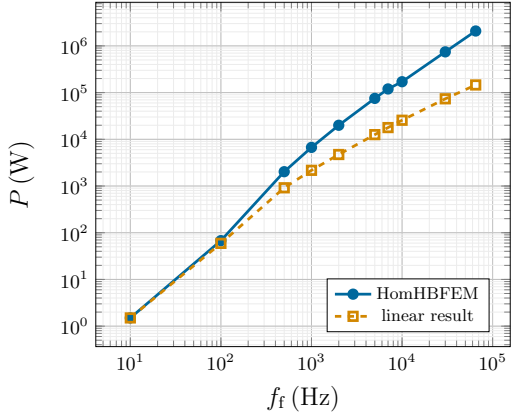
\begin{figure}[H]
	\centering
	\begin{tikzpicture}[scale=0.8]
		\begin{loglogaxis}[
			xlabel = {$\ff$\,(\si{\hertz})},
			ylabel ={$P$\,(\si{\watt})},
			xlabel style={font=\large},
			ylabel style={font=\large},
			grid = both,
			minor grid style = {gray!15},
			minor grid style = {gray!15},
			legend style={at={(0.95,0.12)},anchor = east,nodes={scale=0.9, transform shape}},
			%title = \textbf{Eddy Current Losses in the Yoke},
			]
			\addplot[darkblue_two,mark=*,line width=1.3pt] table[x=f(Hz), y=P(W)] {data_fig_23_nl.txt};
			\addlegendentry{HomHBFEM};
			\addplot[darkyellow,dashed,line width=1.3pt,mark=square, mark options={solid}] table[x=f(Hz), y=P(W)] {data_fig_23_lin.txt};
			\addlegendentry{linear result};
		\end{loglogaxis}
	\end{tikzpicture}
	\caption{Eddy current losses in the FC magnet with $\SI{1}{\milli\meter}$ laminations as a function of frequency computed with the HomHBFEM, compared to the linear results.}
	\label{fig:eddy_current_losses_corrector}
\end{figure}

Figure~\ref{fig:eddy_current_losses_corrector} shows the comparison of the eddy current losses in the yoke computed from the linear and nonlinear solvers. Up to roughly $\SI{100}{\hertz}$, the losses are similar for both cases, but already at $\SI{500}{\hertz}$ the nonlinear simulation gives us much higher power loss. Note that these losses are computed for the same excitation current and applied voltage across the whole frequency range. In reality, the current will decrease significantly as the frequency is increased. Nonetheless, we see from these results that the losses are higher than previously expected based on the linear simulation results. 

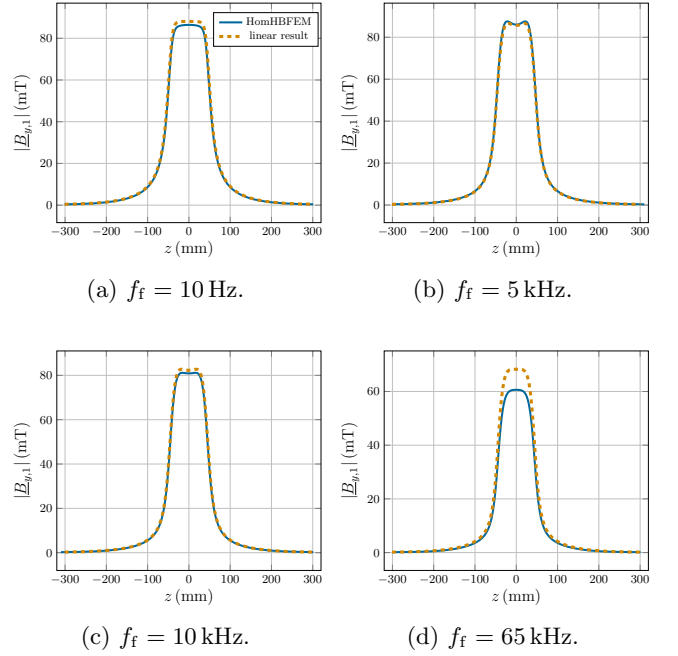
\begin{figure}[t] 
	\centering
	% First row of figures
	\begin{subfigure}{0.23\textwidth}
		\centering
		\begin{tikzpicture}[scale=0.51]
			\begin{axis}[
				xlabel = {$z\,(\si{\milli\meter})$},
				xlabel style={ yshift = 5pt,font=\large},
				ylabel= {$|\underline{B}_{y,1}| \,(\si{\milli\tesla})$},
				ylabel style={ yshift = -10pt,font=\large},
				%title = \textbf{Vertical Field Along the Axis, $f = \SI{10}{\hertz}$},
				grid = both,
				minor grid style = {gray!15},
				legend style={at={(0.99,0.88)},anchor = east,nodes={scale=0.9, transform shape}},
				x coord trafo/.code={\pgfmathparse{#1*1000}\pgfmathresult},
				y coord trafo/.code={\pgfmathparse{#1*1000}\pgfmathresult},
				xmin = -320e-03,
				xmax = 320e-03,
				]
				\addplot[line width = 1.5pt, color=darkblue_two] table {data_fig_24a_nl.txt};
				\addlegendentry{HomHBFEM};
				\addplot[line width = 2.25pt, color=darkyellow,dashed] table {data_fig_24a_lin.txt};
				\addlegendentry{linear result};
			\end{axis}
		\end{tikzpicture}
		\caption{$\ff = \SI{10}{\hertz}$.}
		\label{fig:field_lin_vs_nl_10Hz_03mm_lam}
	\end{subfigure}%
	\hspace{0.1cm} % Adjust the space between the minipages
	\begin{subfigure}{0.23\textwidth}
		\centering
		\begin{tikzpicture}[scale=0.51]
			\begin{axis}[
				xlabel = {$z\,(\si{\milli\meter})$},
				xlabel style={ yshift = 5pt,font=\large},
				ylabel= {$|\underline{B}_{y,1}| \,(\si{\milli\tesla})$},
				ylabel style={ yshift = -10pt,font=\large},
				%title = \textbf{Vertical Field Along the Axis, $f = \SI{5}{\kilo\hertz}$},
				grid = both,
				minor grid style = {gray!15},
				legend style={at={(0.99,0.9)},anchor = east,nodes={scale=0.8, transform shape}},
				x coord trafo/.code={\pgfmathparse{#1*1000}\pgfmathresult},
				y coord trafo/.code={\pgfmathparse{#1*1000}\pgfmathresult},
				xmin = -320e-03,
				xmax = 320e-03,
				]
				\addplot[line width = 1.5pt, color=darkblue_two] table {data_fig_24b_nl.txt};
				%\addlegendentry{Nonlinear};
				\addplot[line width = 2.25pt, color=darkyellow,dashed] table {data_fig_24b_lin.txt};
				%\addlegendentry{Linear};
			\end{axis}
		\end{tikzpicture}
		\caption{$\ff = \SI{5}{\kilo\hertz}$.}
		\label{fig:field_lin_vs_nl_5000Hz_03mm_lam}
	\end{subfigure}
	
	\vspace{0.5cm} % Adjust space between rows if needed
	
	% Second row of figures
	\begin{subfigure}{0.23\textwidth}
		\centering
		\begin{tikzpicture}[scale=0.51]
			\begin{axis}[
				xlabel = {$z\,(\si{\milli\meter})$},
				xlabel style={ yshift = 5pt,font=\large},
				ylabel= {$|\underline{B}_{y,1}| \,(\si{\milli\tesla})$},
				ylabel style={ yshift = -10pt,font=\large},
				%title = \textbf{Vertical Field Along the Axis, $f = \SI{10}{\kilo\hertz}$},
				grid = both,
				minor grid style = {gray!15},
				legend style={at={(0.99,0.9)},anchor = east,nodes={scale=0.8, transform shape}},
				x coord trafo/.code={\pgfmathparse{#1*1000}\pgfmathresult},
				y coord trafo/.code={\pgfmathparse{#1*1000}\pgfmathresult},
				xmin = -320e-03,
				xmax = 320e-03,
				]
				\addplot[line width = 1.5pt, color=darkblue_two] table {data_fig_24c_nl.txt};
				% \addlegendentry{Nonlinear};
				\addplot[line width = 2.25pt, color=darkyellow,dashed] table {data_fig_24c_lin.txt};
				%\addlegendentry{Linear};
			\end{axis}
		\end{tikzpicture}
		\caption{$\ff = \SI{10}{\kilo\hertz}$.}
		\label{fig:field_lin_vs_nl_10000Hz_03mm_lam}
	\end{subfigure}%
	\hspace{0.1cm} % Adjust the space between the minipages
	\begin{subfigure}{0.23\textwidth}
		\centering
		\begin{tikzpicture}[scale=0.51]
			\begin{axis}[
				xlabel = {$z\,(\si{\milli\meter})$},
				xlabel style={ yshift = 5pt,font=\large},
				ylabel= {$|\underline{B}_{y,1}| \,(\si{\milli\tesla})$},
				ylabel style={ yshift = -10pt,font=\large},
				%title = \textbf{Vertical Field Along the Axis, $f = \SI{65}{\kilo\hertz}$},
				grid = both,
				minor grid style = {gray!15},
				legend style={at={(0.99,0.9)},anchor = east,nodes={scale=0.8, transform shape}},
				x coord trafo/.code={\pgfmathparse{#1*1000}\pgfmathresult},
				y coord trafo/.code={\pgfmathparse{#1*1000}\pgfmathresult},
				xmin = -320e-03,
				xmax = 320e-03,
				]
				\addplot[line width = 1.5pt, color=darkblue_two] table {data_fig_24d_nl.txt};
				% \addlegendentry{Nonlinear};
				\addplot[line width = 2.25pt, color=darkyellow,dashed] table {data_fig_24d_lin.txt};
				% \addlegendentry{Linear};
			\end{axis}
		\end{tikzpicture}
		\caption{$\ff = \SI{65}{\kilo\hertz}$.}
		\label{fig:field_lin_vs_nl_65000Hz_03mm_lam}
	\end{subfigure}
	\caption{Magnitude of the dipole component of the magnetic flux density's first harmonic along the axis of the FC magnet computed with the HomHBFEM for $\SI{0.3}{\milli\meter}$ laminations at four fundamental frequencies of the source current, compared to the linear results.}
	\label{fig:vertical_field_along_axis_03mm_lam}
\end{figure}

\begin{figure}[b]
	\centering
	\begin{tikzpicture}[scale=0.8]
		\begin{loglogaxis}[
			xlabel = {$\ff$\,(\si{\hertz})},
			ylabel ={$P$\,(\si{\watt})},
			xlabel style={ font=\large},
			xlabel style={ font=\large},
			grid = both,
			minor grid style = {gray!15},
			minor grid style = {gray!15},
			legend style={at={(0.95,0.12)},anchor = east,nodes={scale=0.9, transform shape}},
			%title = \textbf{Eddy Current Losses in the Yoke},
			]
			\addplot[darkblue_two,mark=*,line width=1.3pt] table[x=f(Hz), y=P(W)] {data_fig_25_nl.txt};
			\addlegendentry{HomHBFEM};
			\addplot[darkyellow,dashed,line width=1.3pt,mark=square, mark options={solid}] table[x=f(Hz), y=P(W)] {data_fig_25_lin.txt};
			\addlegendentry{linear result};
		\end{loglogaxis}
	\end{tikzpicture}
	\caption{Eddy current losses in the FC magnet with $\SI{0.3}{\milli\meter}$ laminations as a function of frequency computed with the HomHBFEM, compared to the linear results.}
	\label{fig:eddy_current_losses_corrector_03mm_lam}
\end{figure}
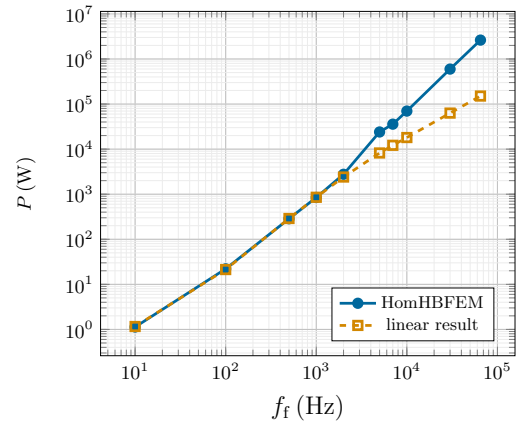

Next, we investigate the magnet model with a lamination thickness of $\SI{0.3}{\milli\meter}$, as intended for series production. As we will see, this change in the lamination thickness significantly decreases the effect of the $B$-$H$ curve's nonlinearity on the behavior of the magnet. Figure ~\ref{fig:vertical_field_along_axis_03mm_lam} shows the results for the longitudinal field profiles with $d = \SI{0.3}{\milli\meter}$ and Fig.~\ref{fig:eddy_current_losses_corrector_03mm_lam} shows the eddy current losses. 

Regarding the longitudinal field profiles, we observe differences between the linear and the nonlinear simulation occuring only at frequencies significantly above $\SI{10}{\kilo\hertz}$, whereas before, with $d = \SI{1}{\milli\meter}$, the differences started to become apparent already between $\SI{1}{\kilo\hertz}$ and $\SI{5}{\kilo\hertz}$. Also with regard to the eddy current losses, the frequency range in which the nonlinear results agree with the linear ones is now broadened. We observe only minor deviations up to $\SI{2}{\kilo\hertz}$, whereas with the thicker lamination, differences are apparent already at $f =\SI{500}{\hertz}$. 

All in all, we observe that the change from $\SI{1}{\milli\meter}$ to $\SI{0.3}{\milli\meter}$ laminations has pushed the effect of the nonlinearity to higher frequencies, which is very beneficial for the design of the feedback control. This phenomenon can be explained from theory as follows. As the frequency is increased, the skin effect leads to a growing difference between the magnetic flux density in the center of each lamination and the magnetic flux density close to its surface. Hence, the reluctivity, which results from these magnetic flux densities and the nonlinear $B$-$H$ curve, will become more and more non-uniform across the laminations. Therefore, as the frequency is increased, the assumption of a constant reluctivity, which is at the core of the linear simulation, becomes less and less valid. That explains why in general, the nonlinearity becomes more relevant at higher frequencies. Since the intensity of the skin effect is not determined by the frequency alone but rather by the ratio $d/ \delta \propto d\sqrt{\omega}$, it also explains why a thinner lamination pushes the effect of the nonlinearity to higher frequencies.

If the magnetic flux density in the laminations is sufficiently small, as is the case for the FC magnet of \mbox{PETRA IV}, then it is clear from the explanation above that the described effect will only occur if the $B$-$H$ curve features a Rayleigh region, i.e., if $B$ does not scale linearly with $H$ for $H \rightarrow 0$, but rather quadratically~\cite{cullity_2009}. Otherwise, the change of the magnetic flux density across each lamination would not translate into a non-uniform reluctivity. This is precisely what we find in the simulation with the HomHBFEM as well. For instance, if we use an adaptation of the original $B$-$H$ curve which eliminates the Rayleigh region, as indicated in Fig.~\ref{fig:bh_curve_powercore_no_rayleigh}, we observe almost no effect of the nonlinearity on the magnetic flux density along the longitudinal axis, even at the higher frequencies and with the $\SI{1}{\milli\meter}$ lamination. This is shown in Fig.~\ref{fig:field_lin_vs_nl_1mm_lam_no_rayleigh}. Hence, we conclude that, indeed, the effect of the nonlinearity in the FC magnets for \mbox{PETRA IV} can be drastically reduced by choosing a material with a less pronounced Rayleigh region. Additionally, we remark that using the adapted $B$-$H$ curve instead of the original one leads to higher magnetic flux densities in the center of the aperture (compare Figs.~\ref{fig:field_lin_vs_nl_1mm_lam_no_rayleigh} and \ref{fig:vertical_field_along_axis}), since the adaptation corresponds to higher permeabilities in the yoke.
\begin{figure}[h]
	\centering
	\resizebox{0.7\linewidth}{!}{\input{fig_26.tex}}
	\caption{$B$-$H$ curve of powercore\textsuperscript{\textregistered} $1400$-$100\text{AP}$ compared to the adapted \mbox{$B$-$H$ curve} without Rayleigh region.}
	\label{fig:bh_curve_powercore_no_rayleigh}
\end{figure}
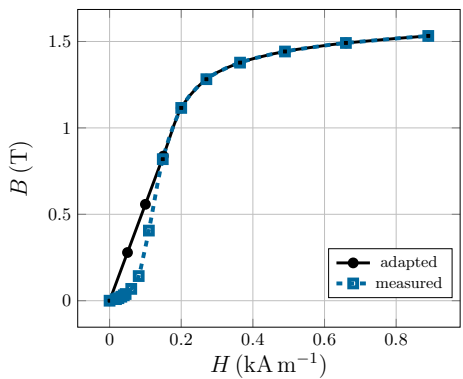
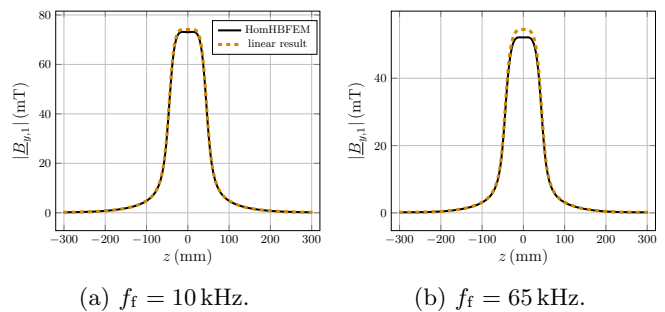
\begin{figure}[h]
	\centering
	% First subfigure
	\begin{subfigure}[b]{0.23\textwidth}
		\centering
		\begin{tikzpicture}[scale=0.51]
			\begin{axis}[
				xlabel = {$z\,(\si{\milli\meter})$},
				xlabel style={ yshift = 5pt,font=\large},
				ylabel= {$|\underline{B}_{y,1}| \,(\si{\milli\tesla})$},
				ylabel style={ yshift = -10pt,font=\large},
				grid = both,
				minor grid style = {gray!15},
				legend style={at={(0.99,0.88)},anchor = east,nodes={scale=0.9, transform shape}},
				x coord trafo/.code={\pgfmathparse{#1*1000}\pgfmathresult},
				y coord trafo/.code={\pgfmathparse{#1*1000}\pgfmathresult},
				xmin = -320e-03,
				xmax = 320e-03,
				]
				\addplot[line width = 1.5pt, color=black] table {data_fig_27a_nl.txt};
				\addlegendentry{HomHBFEM};
				\addplot[line width = 2.25pt, color=darkyellow,dashed] table {data_fig_27a_lin.txt};
				\addlegendentry{linear result};
			\end{axis}
		\end{tikzpicture}
		\caption{$\ff = \SI{10}{\kilo\hertz}$.}
		\label{fig:field_lin_vs_nl_10000Hz_1mm_lam_no_rayleigh}
	\end{subfigure}
	\hspace{0.1cm}
	\begin{subfigure}[b]{0.23\textwidth}
		\centering
		\begin{tikzpicture}[scale=0.51]
			\begin{axis}[
				xlabel = {$z\,(\si{\milli\meter})$},
				xlabel style={ yshift = 5pt, font=\large},
				ylabel= {$|\underline{B}_{y,1}| \,(\si{\milli\tesla})$},
				ylabel style={ yshift = -10pt,font=\large},
				grid = both,
				minor grid style = {gray!15},
				legend style={at={(0.99,0.9)},anchor = east,nodes={scale=0.8, transform shape}},
				x coord trafo/.code={\pgfmathparse{#1*1000}\pgfmathresult},
				y coord trafo/.code={\pgfmathparse{#1*1000}\pgfmathresult},
				xmin = -320e-03,
				xmax = 320e-03,
				]
				\addplot[line width = 1.5pt, color=black] table {data_fig_27b_nl.txt};
				%\addlegendentry{HomHBFEM};
				\addplot[line width = 2.25pt, color=darkyellow,dashed] table {data_fig_27b_lin.txt};
				% \addlegendentry{linear};
			\end{axis}
		\end{tikzpicture}
		\caption{$\ff = \SI{65}{\kilo\hertz}$.}
		\label{fig:field_lin_vs_nl_65000Hz_1mm_lam_no_rayleigh}
	\end{subfigure}
	\caption{Magnitude of the dipole component of the magnetic flux density's first harmonic along the axis of the FC magnet computed with the HomHBFEM for $\SI{1}{\milli\meter}$ laminations and the adapted $B$-$H$ curve without Rayleigh region, compared to the linear results.}
	\label{fig:field_lin_vs_nl_1mm_lam_no_rayleigh}
\end{figure}

\section{Conclusion}\label{sec:conclusion}
In this paper, we have shown how an existing frequency domain based homogenization technique for lamination stacks can be combined with the harmonic balance finite element method (HBFEM) to consider nonlinear \mbox{$B$-$H$ curves}. The resulting method, HomHBFEM, is designed to conduct nonlinear eddy current simulations efficiently, without time-stepping, and with a relatively coarse mesh. In particular, the mesh does not have to resolve the individual lamination sheets nor the skin depth therein for the simulation results to converge. As a result, the HomHBFEM can drastically reduce simulation times at elevated frequencies compared to a transient simulation. This has been demonstrated using two different models for verification, a laminated inductor and a C-shaped dipole magnet. In particular, the verification studies confirmed that the HomHBFEM can be employed to determine the eddy current losses and the field in a magnet's aperture in the kilohertz range. In this context, it is important to highlight that as the level of saturation increases, a higher number of harmonics should be included in the analysis to maintain accuracy. For the FC magnets of \mbox{PETRA IV} at \mbox{DESY}, the method has enabled us to conduct nonlinear simulations over a broad frequency range up to $\SI{65}{\kilo\hertz}$, which was impossible with the available computational resources when using a standard 3D FE simulation. This investigation of the FC magnets with the HomHBFEM has yielded valuable insights regarding the impact of the nonlinearity on the dynamic behavior of the magnets and how it can be mitigated. 
Hence, the HomHBFEM approach is also of interest for other applications with laminated ferromagnetic yokes or cores, e.g., other types of fast accelerator magnets, transformers, and electrical machines.

\begin{acknowledgments}
We acknowledge the support from Deutsches Elektronen-Synchrotron DESY and the German Research Foundation (DFG) - GRK 2128 "AccelencE". Furthermore, we would like to thank Nicolas Marsic for the fruitful discussions on GetDP implementation issues. 
\end{acknowledgments}
\appendix*
\section{Probed Locations in the Laminated Inductor Model}\label{sec:appendix}
Table~\ref{tab:locations} gives the coordinates of the locations that were probed for the plots of the magnetic flux densities in Sec.~\ref{sec:toy_model}. The ferromagnetic core of the laminated inductor model extends $\SI{5}{\milli\meter}$ in $x$-direction, $\SI{12.5}{\milli\meter}$ in $y$-direction and $\SI{5.07}{\milli\meter}$ in $z$-direction. The coordinate system is aligned with one of the core's corners and, using the symmetry planes of the model, we simulated only the octant in the opposite corner.
\begin{table}[H]
	\centering
	\caption{Coordinates of probed locations.}\label{tab:locations}
	\begin{tabular}{cccc} % 'lcc' sets alignment: left, center, center for the three columns
		\toprule
		\toprule
		\textbf{Location} & $x (\si{\milli\meter})$ & $y (\si{\milli\meter})$ & $z (\si{\milli\meter})$ \\
		\midrule
		$1$ & $3.125$ & $7.878$ & $3.893$ \\
		$2$ & $2.813$ & $10.246$ & $3.623$ \\
		$3$ & $4.061$ & $11.569$ & $2.813$ \\
		$4$ & $3.100$ & $8.951$ & $4.588$\\
		$5$ & $4.716$ & $11.569$ & $4.0333$\\
		$6$ & $3.344$ & $8.291$ & $3.056$\\
		$7$ & $4.770$ & $8.297$ & $4.696$ \\
		$8$ & $4.566$ & $7.787$ & $4.643$ \\
		\bottomrule
		\bottomrule
	\end{tabular}
\end{table}

%\nocite{*}

\bibliography{Paper_HomHBFEM_1_aps}% Produces the bibliography via BibTeX.

\end{document}

%% file: fig_1.tex
	
	\begin{tikzpicture}
		% Define parameters
		\def\width{5}        % Width of brick
		\def\height{6}       % Height of brick (same for all layers)
		\def\blueDepth{1}    % Depth of blue layers (full thickness)
		\def\orangeDepth{0.3} % Depth of orange layers (half the thickness of blue)
		
		% Initialize the starting z-coordinate
		\def\zcoord{0}
		
		% Blue layer 1 (full thickness)
		\fill[darkblue_two!100] (0, 0, \zcoord) -- ++(\width, 0, 0) -- ++(0, \height, 0) -- ++(-\width, 0, 0) -- cycle;
		\fill[darkblue_two!70] (0, \height, \zcoord) -- ++(\width, 0, 0) -- ++(0, 0, \blueDepth) -- ++(-\width, 0, 0) -- cycle;
		\fill[darkblue_two!60] (\width, 0, \zcoord) -- ++(0, \height, 0) -- ++(0, 0, \blueDepth) -- ++(0, -\height, 0) -- cycle;
		\pgfmathsetmacro{\zcoord}{\zcoord + \blueDepth} % Update z-coordinate
		
		% Orange layer 1 (half thickness)
		\fill[darkred!100] (0, 0, \zcoord) -- ++(\width, 0, 0) -- ++(0, \height, 0) -- ++(-\width, 0, 0) -- cycle;
		\fill[darkred!70] (0, \height, \zcoord) -- ++(\width, 0, 0) -- ++(0, 0, \orangeDepth) -- ++(-\width, 0, 0) -- cycle;
		\fill[darkred!60] (\width, 0, \zcoord) -- ++(0, \height, 0) -- ++(0, 0, \orangeDepth) -- ++(0, -\height, 0) -- cycle;
		\pgfmathsetmacro{\zcoord}{\zcoord + \orangeDepth} % Update z-coordinate
		
		% Blue layer 2 (full thickness)
		\fill[darkblue_two!100] (0, 0, \zcoord) -- ++(\width, 0, 0) -- ++(0, \height, 0) -- ++(-\width, 0, 0) -- cycle;
		\fill[darkblue_two!70] (0, \height, \zcoord) -- ++(\width, 0, 0) -- ++(0, 0, \blueDepth) -- ++(-\width, 0, 0) -- cycle;
		\fill[darkblue_two!60] (\width, 0, \zcoord) -- ++(0, \height, 0) -- ++(0, 0, \blueDepth) -- ++(0, -\height, 0) -- cycle;
		\pgfmathsetmacro{\zcoord}{\zcoord + \blueDepth} % Update z-coordinate
		
		% Orange layer 2 (half thickness)
		\fill[darkred!100] (0, 0, \zcoord) -- ++(\width, 0, 0) -- ++(0, \height, 0) -- ++(-\width, 0, 0) -- cycle;
		\fill[darkred!70] (0, \height, \zcoord) -- ++(\width, 0, 0) -- ++(0, 0, \orangeDepth) -- ++(-\width, 0, 0) -- cycle;
		\fill[darkred!60] (\width, 0, \zcoord) -- ++(0, \height, 0) -- ++(0, 0, \orangeDepth) -- ++(0, -\height, 0) -- cycle;
		\pgfmathsetmacro{\zcoord}{\zcoord + \orangeDepth} % Update z-coordinate
		
		% Blue layer 3 (full thickness)
		\fill[darkblue_two!100] (0, 0, \zcoord) -- ++(\width, 0, 0) -- ++(0, \height, 0) -- ++(-\width, 0, 0) -- cycle;
		\fill[darkblue_two!70] (0, \height, \zcoord) -- ++(\width, 0, 0) -- ++(0, 0, \blueDepth) -- ++(-\width, 0, 0) -- cycle;
		\fill[darkblue_two!60] (\width, 0, \zcoord) -- ++(0, \height, 0) -- ++(0, 0, \blueDepth) -- ++(0, -\height, 0) -- cycle;
		\pgfmathsetmacro{\zcoord}{\zcoord + \blueDepth} % Update z-coordinate
		
		% Orange layer 3 (half thickness)
		\fill[darkred!100] (0, 0, \zcoord) -- ++(\width, 0, 0) -- ++(0, \height, 0) -- ++(-\width, 0, 0) -- cycle;
		\fill[darkred!70] (0, \height, \zcoord) -- ++(\width, 0, 0) -- ++(0, 0, \orangeDepth) -- ++(-\width, 0, 0) -- cycle;
		\fill[darkred!60] (\width, 0, \zcoord) -- ++(0, \height, 0) -- ++(0, 0, \orangeDepth) -- ++(0, -\height, 0) -- cycle;
		\pgfmathsetmacro{\zcoord}{\zcoord + \orangeDepth} % Update z-coordinate
		
		% Blue layer 4 (full thickness)
		\fill[darkblue_two!100] (0, 0, \zcoord) -- ++(\width, 0, 0) -- ++(0, \height, 0) -- ++(-\width, 0, 0) -- cycle;
		\fill[darkblue_two!70] (0, \height, \zcoord) -- ++(\width, 0, 0) -- ++(0, 0, \blueDepth) -- ++(-\width, 0, 0) -- cycle;
		\fill[darkblue_two!60] (\width, 0, \zcoord) -- ++(0, \height, 0) -- ++(0, 0, \blueDepth) -- ++(0, -\height, 0) -- cycle;
		
		\pgfmathsetmacro{\zcoord}{\zcoord + \blueDepth} % Update z-coordinate
		\fill[darkblue_two!100] (0, 0, \zcoord) -- ++(\width, 0, 0) -- ++(0, \height, 0) -- ++(-\width, 0, 0) -- cycle;
		
		% coordinate system 
		\draw[->, line width= 2pt] (-3.1, -2) -- (-3.1, -0.8);
		\node at (-3.1, -0.8) [anchor=east] {\huge $y$};
		
		\draw[->, line width= 2pt] (-3.1, -2) -- (-1.9, -2);
		\node at (-1.9, -2.1) [anchor=north] {\huge $x$};

		\draw[line width= 2pt] (-3.1,-2) circle [radius=0.2cm];
		\filldraw (-3.1,-2) circle (2.5pt);
		\node at (-3.2, -2.35) [anchor=east] {\huge $z$};
		
		% material labels
		\draw[->,line width = 2pt] (-0.5,6.5)--(0,5.8);
		\node at (-1.5,6.8) [anchor=west] {\huge $\nu_{\mathrm{c}}, \sigma_{\mathrm{c}}  $};
		
		\draw[->, line width = 2pt] (-1.3,5.75)--(-0.7,5.05);
		\node at (-2.9,5.95) [anchor=west] {\huge $\nu_{\mathrm{ins}}$, $\sigma_{\mathrm{ins}}$};

		% homogenized structure
		\def\zcoord{0}
		\def\xcoord{9}
		\def\Depth{4.9}    % Depth of blue layers (full thickness)
		% Blue layer 1 (full thickness)
		\fill[darkblue_two!100] (\xcoord, 0, \zcoord) -- ++(\width, 0, 0) -- ++(0, \height, 0) -- ++(-\width, 0, 0) -- cycle;
		\fill[darkblue_two!70] (\xcoord, \height, \zcoord) -- ++(\width, 0, 0) -- ++(0, 0, \Depth) -- ++(-\width, 0, 0) -- cycle;
		\fill[darkblue_two!60] (\xcoord + \width, 0, \zcoord) -- ++(0, \height, 0) -- ++(0, 0, \Depth) -- ++(0, -\height, 0) -- cycle;

		\pgfmathsetmacro{\zcoord}{\zcoord + \Depth} % Update z-coordinate
		\fill[darkblue_two!100] (\xcoord, 0, \zcoord) -- ++(\width, 0, 0) -- ++(0, \height, 0) -- ++(-\width, 0, 0) -- cycle;
		
		\draw[->, line width = 2pt] (5+ \xcoord,6.5)--(4.5 + \xcoord,5.8);
		\node at (4.5+ \xcoord,7.0) [anchor=west] {\huge $\underline{\overline{\overline{\nu}}}, \overline{\overline{\sigma}}$};
		
		% arc pointing from full to hom. model
		
		\draw[->, line width = 2pt] (\xcoord -5, 5.8) arc [start angle = 180,end angle =0, x radius = 2.9, y radius =1]; 
		\node at (\xcoord - 2,7) [anchor = south] {\huge homogenization technique};
		
	\end{tikzpicture}
	

%% file: fig_4.tex
	\begin{tikzpicture}[scale = 0.5]
		\begin{axis}[
			xlabel = {$H\,(\si{\kilo\ampere/\meter})$},
			xlabel style={ yshift = 0pt, font=\large},
			ylabel= {$B\,(\si{\tesla})$},
			ylabel style={ yshift = -15pt, font=\large},
			grid = both,
			minor grid style = {gray!15},
			minor grid style = {gray!15},
			legend style={at={(0.05,0.15)},anchor = west,nodes={scale=1.0, transform shape}},
			x filter/.code={\pgfmathparse{0.001*\pgfmathresult}},
			]
			\addplot[line width = 1.5pt, color=darkblue_two] table {data_fig_4.txt};
		\end{axis}
	\end{tikzpicture}

%% file: fig_14.tex
	\begin{tikzpicture}[scale=0.64, every node/.style={transform shape}]
		% Yoke
		\coordinate (A) at (0,0);
		\coordinate (B) at (5,0); 
		\coordinate (C) at (5,3.5);
		\coordinate (D) at (3,3.5);
		\coordinate (E) at (3,2);
		\coordinate (F) at (2,2);
		\coordinate (G) at (2,7.5);
		\coordinate (H) at (3,7.5);
		\coordinate (I) at (3,6);
		\coordinate (J) at (5,6);
		\coordinate (K) at (5,9.5);
		\coordinate (L) at (0,9.5);
		\draw [draw,ultra thick, darkblue_two] (A)--(B)--(C)--(D)--(E)--(F)--(G)--(H)--(I)--(J)--(K)--(L)--cycle; 
		% Coil lower left
		\coordinate (ALowCoilLeft)at (2.1,2.35);
		\coordinate (BLowCoilLeft)at (2.9,2.35);
		\coordinate (CLowCoilLeft)at (2.9,3.15);
		\coordinate (DLowCoilLeft)at (2.1,3.15);	
		\draw [ultra thick, darkred] (ALowCoilLeft)--(BLowCoilLeft);
		\draw [ultra thick, darkred] (CLowCoilLeft)--(DLowCoilLeft);
		\draw [ultra thick, darkred] (ALowCoilLeft)--(CLowCoilLeft);
		\draw [ultra thick, darkred] (BLowCoilLeft)--(DLowCoilLeft);
		\draw [ultra thick, darkred] (ALowCoilLeft)--(DLowCoilLeft);
		\draw [ultra thick, darkred] (BLowCoilLeft)--(CLowCoilLeft);
		% Coil lower right
		\coordinate (ALowCoilRight)at (5.1,2.35);
		\coordinate (BLowCoilRight)at (5.9,2.35);
		\coordinate (CLowCoilRight)at (5.9,3.15);
		\coordinate (DLowCoilRight)at (5.1,3.15);
		\coordinate(LowCoilRightCenter) at (5.5,2.75);
		\draw [ultra thick, darkred] (ALowCoilRight)--(BLowCoilRight);
		\draw [ultra thick, darkred] (BLowCoilRight)--(CLowCoilRight);
		\draw [ultra thick, darkred] (CLowCoilRight)--(DLowCoilRight);
		\draw [ultra thick, darkred] (DLowCoilRight)--(ALowCoilRight);
		\filldraw[darkred] (LowCoilRightCenter) circle (2pt);
		% Coil upper left
		\coordinate (AHighCoilLeft)at (2.1,6.35);
		\coordinate (BHighCoilLeft)at (2.9,6.35);
		\coordinate (CHighCoilLeft)at (2.9,7.15);
		\coordinate (DHighCoilLeft)at (2.1,7.15);	
		\draw [ultra thick, darkred] (AHighCoilLeft)--(BHighCoilLeft);
		\draw [ultra thick, darkred] (CHighCoilLeft)--(DHighCoilLeft);
		\draw [ultra thick, darkred] (AHighCoilLeft)--(CHighCoilLeft);
		\draw [ultra thick, darkred] (BHighCoilLeft)--(DHighCoilLeft);
		\draw [ultra thick, darkred] (AHighCoilLeft)--(DHighCoilLeft);
		\draw [ultra thick, darkred] (BHighCoilLeft)--(CHighCoilLeft);
		% Coil upper right1
		\coordinate (AHighCoilRight)at (5.1,6.35);
		\coordinate (BHighCoilRight)at (5.9,6.35);
		\coordinate (CHighCoilRight)at (5.9,7.15);
		\coordinate (DHighCoilRight)at (5.1,7.15);
		\coordinate (HighCoilRightCenter) at (5.5,6.75);
		\draw [ultra thick, darkred] (AHighCoilRight)--(BHighCoilRight);
		\draw [ultra thick, darkred] (BHighCoilRight)--(CHighCoilRight);
		\draw [ultra thick, darkred] (CHighCoilRight)--(DHighCoilRight);
		\draw [ultra thick, darkred] (DHighCoilRight)--(AHighCoilRight);
		\filldraw[darkred] (HighCoilRightCenter) circle (2pt);
		% Beampipe
		%\coordinate (BPCenter) at (4,4.75);
		%\def\rin{1.0};
		%\def\rout{1.05};
		%\draw[ultra thick] (BPCenter) circle (\rin);
		%\draw[ultra thick] (BPCenter) circle (\rout);
		% dimensions
		\draw[arrows = <->,thick,color = darkblue_two] (3.0,6.35) -- (5.0,6.35) node[midway,above] {\Large $\SI{20}{\milli\meter}$};
		\draw[arrows = <->,thick,color = darkblue_two] (2.0,6.35) -- (0,6.35) node[midway,above] {\Large $\SI{20}{\milli\meter}$};
		\draw[arrows = <->,thick,color = darkblue_two] (2,7.75) -- (3,7.75) node[midway,above, ,yshift=2pt] {\Large $\SI{10}{\milli\meter}$};
		\draw[arrows = <->,thick,color = darkblue_two] (-0.25,0) -- (-0.25,9.5) node[midway,above,rotate = 90] {\Large $\SI{95}{\milli\meter}$};
		\draw[arrows = <->,thick,color = darkblue_two] (5.25,3.5) -- (5.25,6) node[midway,below,rotate = 90] {\Large $\SI{25}{\milli\meter}$};
		%\draw[arrows = ->,thick,color = red] (4,4.75) -- (5,4.75) node[midway,above,,scale = 0.8] {$\SI{10}{\milli\meter}$};
		%\draw[arrows = ->,thick,color = red] (3.4,4.75) -- (3,4.75) node[above,scale = 0.9] {};
		%\draw[arrows = ->,thick,color = red] (2.6,4.75) -- (2.95,4.75) node[above,scale = 0.5] {};
		%\node[thick,red,scale = 0.5] (BP_Thickness) at (2.6,5){$\SI{0.5}{\milli\meter}$};
	\end{tikzpicture}

%% file: fig_21a.tex
        \begin{tikzpicture}[scale=0.5]
            \begin{axis}[
                xlabel = {$H\,(\si{\kilo\ampere\per\meter})$},
                xlabel style={font=\large},
                ylabel= {$B\,(\si{\tesla})$},
                 ylabel style={ font=\large},
                grid = both,
                minor grid style = {gray!15},
                legend style={at={(0.99,0.9)},anchor = east,nodes={scale=0.8, transform shape}},
                x coord trafo/.code={\pgfmathparse{#1*0.001}\pgfmathresult},
            ]
                 \addplot[line width = 1.5pt, color=darkblue_two, only marks, mark=square] table {data_fig_21a_1.txt};
                \addplot[line width = 1.5pt, color=darkblue_two,dashed] table {data_fig_21a_2.txt};
            \end{axis}
        \end{tikzpicture}

%% file: fig_21b.tex
	\begin{tikzpicture}[scale=0.5]
		\begin{axis}[
			xlabel = {$H\,(\si{\ampere/\meter})$},
			xlabel style={ yshift = 0pt, font=\large},
			yticklabel style={
				/pgf/number format/1000 sep={}
			},
			ylabel= {$\mu_{\mathrm{r}}$},
			ylabel style={font=\large},
			grid = both,
			minor grid style = {gray!15},
			major grid style = {gray!25},
			legend style={at={(0.05,0.15)},anchor = west,nodes={scale=1.0, transform shape}},
			xmode=log,
			ymax = 4800,
			xmin = 10,
			xmax = 3e+04,
			y filter/.code={\pgfmathparse{795774.7155*\pgfmathresult}},
			]
			 \addplot[line width = 1.5pt, color=darkblue_two, only marks, mark=square] table {data_fig_21b_1.txt};
			\addplot[line width = 1.5pt, color=darkblue_two,dashed] table {data_fig_21b_2.txt};
		\end{axis}
	\end{tikzpicture}

%% file: fig_26.tex
        \begin{tikzpicture}[scale=0.45]
            \begin{axis}[
                xlabel = {$H\,(\si{\kilo\ampere\per\meter})$},
                xlabel style={ yshift = 5pt, font=\large},
                ylabel= {$B \,(\si{\tesla})$},
                ylabel style={ yshift = -6pt, font=\large},
                %title = \textbf{$BH$-curve of Powercore 1400AP},
                grid = both,
                minor grid style = {gray!15},
                legend style={at={(0.98,0.17)},anchor = east,nodes={scale=0.9, transform shape}},
                x coord trafo/.code={\pgfmathparse{#1*0.001}\pgfmathresult},
            ]
            	 \addplot[line width = 1.5pt, color=black, smooth, mark=*, mark options={solid}, restrict expr to domain={\coordindex}{0:9}] table {data_fig_26_1.txt};   
            	 \addlegendentry{adapted};
                \addplot[line width = 2pt, color=darkblue_two,dashed, smooth,  mark=square, mark options={solid},restrict expr to domain={\coordindex}{0:15}] table {data_fig_26_2.txt};
                 \addlegendentry{measured};
                    
            \end{axis}
        \end{tikzpicture}